\documentclass[conference]{IEEEtran}
\IEEEoverridecommandlockouts
\usepackage[utf8]{inputenc}
\usepackage[english]{babel}
\usepackage{url}
\usepackage{placeins}[section]
\usepackage{flushend}
\usepackage{amsmath,amssymb,amsfonts,amsthm,bm,bbm}
\usepackage{paralist}
\usepackage{fancybox}
\usepackage{csquotes}
\usepackage{booktabs}
\usepackage{threeparttable}

\DeclareMathOperator{\EX}{\mathbb{E}}
\newcommand{\proto}{Interlude}
\usepackage{algorithm}
\usepackage{algpseudocode}
\usepackage{graphicx}
\usepackage{textcomp}
\usepackage[dvipsnames,table]{xcolor}
\usepackage{pifont}
\newcommand{\cmark}{{\color{OliveGreen}\ding{51}}}%
\newcommand{\xmark}{{\color{BrickRed}\ding{55}}}%
\usepackage{multirow,array}

\newtheorem{theorem}{Theorem}[section]

\newtheorem{lemma}[theorem]{Lemma}
\theoremstyle{definition}
\newtheorem{definition}{Definition}[section]

\newcommand{\hl}[1]{#1}
\newcommand{\st}[1]{}

\definecolor{Gray}{gray}{0.95}
\def\thesubsubsectiondis{\unskip\arabic{subsubsection})}
\begin{document}

\title{Interlude: Balancing Chaos And Harmony For Fair and Fast Blockchains}
\author{\IEEEauthorblockN{Anurag Jain}
\IEEEauthorblockA{\textit{Machine Learning Lab} \\
\textit{International Institute of Information Technology}\\
Hyderabad, India \\
\texttt{anurag.jain@research.iiit.ac.in}}
\and
\IEEEauthorblockN{Kannan Srinathan}
\IEEEauthorblockA{\textit{Center for Security, Theory \& Algorithmic Research} \\
\textit{International Institute of Information Technology}\\
Hyderabad, India \\
\texttt{srinathan@iiit.ac.in}}
\and
\IEEEauthorblockN{Sujit Gujar}
\IEEEauthorblockA{\textit{Machine Learning Lab} \\
\textit{International Institute of Information Technology}\\
Hyderabad, India \\
\texttt{sujit.gujar@iiit.ac.in}}
}
\maketitle
\begin{abstract}
Blockchains lie at the heart of Bitcoin and other cryptocurrencies that have shown great promise to revolutionize finance and commerce. Although they are gaining increasing popularity, they face technical challenges when it comes to scaling to support greater demand while maintaining their desirable security properties. In an exciting line of recent work, many researchers have proposed various scalable blockchain protocols that demonstrate the potential to solve these challenges. However, many of these protocols come with the assumptions of honest majority and symmetric network access which may not accurately reflect the real world where the participants may be self-interested or rational. Secondly, these works show that their protocol works in an ideal environment where each party has equal access to the network whereas different parties have varying latencies and network speeds. These assumptions may render the protocols susceptible to security threats in the real world, as highlighted by the literature focused on exploring game-theoretic attacks on these protocols.

We propose a scalable blockchain protocol, \emph{\proto}, which comes with the typical security guarantees while focusing on game-theoretic soundness and network fairness. The novelty of \proto\ is that it has a relatively simple design consisting of a sequence of parallel blocks containing disjoint transaction sets that can be mined quickly followed by a series block that is slow to mine and gives the honest parties in the network time to synchronize. Thus, between the chaos of parallel blocks, our blockchain protocol masquerades an interlude moment of harmony in series blocks that synchronize the network.
\end{abstract}

\section{Introduction}
\begin{figure*}[tb]
\centering
\includegraphics[width=\linewidth]{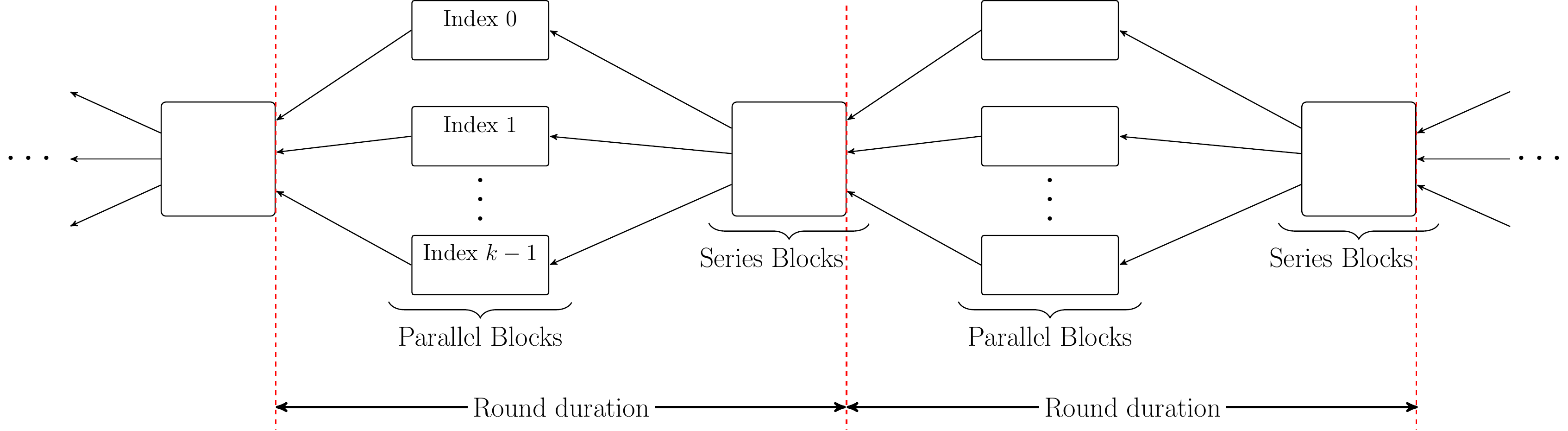}
\caption{An illustration of a chain in \proto\label{fig:main}}
\end{figure*}

Blockchains are heralded as a technological innovation poised to revolutionize finance and commerce. This reputation is drawn from the fact that to date, it is the only technology that can allow two mutually mistrusting entities to exchange financial value without relying on a trusted third party. Moreover, blockchains can also securely store data in a timestamped and tamper-proof manner, enabling them to serve more general applications such as smart contracts.

One of the most common applications of blockchain is to serve as a building block for cryptocurrencies, Bitcoin and Ethereum being well-known examples. Widespread adoption of such cryptocurrencies require blockchain protocols to offer greater throughputs comparable to existing payment systems like the Visa Network. Yet, today's cryptocurrencies are still not a feasible replacement for centralized payment systems like Visa since they lag behind in performance parameters. As of December 2021, Bitcoin's and Ethereum's network processes an average of 3-4 and 10 transactions per second (TPS), respectively.  In contrast, Visa's global payment system handles a reported 1,700 TPS and claims to be capable of handling more than 24,000 TPS \cite{visa_claim}.  Hence, the inability of existing blockchain protocols to scale to higher throughputs is a concern among researchers and protocol designers.
\subsection{Scalability and Security}
An important issue that a blockchain protocol needs to solve is maintaining security against a \emph{byzantine} adversary while scaling the protocol. Garay et al. \cite{cryptoeprint:2014:765} and Kiayias et al. \cite{cryptoeprint:2015:1019} show that existing blockchain protocols such as Bitcoin suffer from a loss of security properties as we scale the system. Consequently, many researchers have attempted to solve this problem by proposing novel blockchain protocols that scale better than Bitcoin \cite{eyal2016bitcoin,cryptoeprint:2013:881,spectre,phantom,li2020decentralized}.
\subsection{Scalability and Game Theory}
Most existing protocols come with with the assumption of honest majority whereas, in the real world participants may be self-interested or rational. Such assumptions may leave them susceptible to incentive-driven deviations and other game-theoretic issues~\cite{10.1007/978-3-030-53356-4_5, DBLP:journals/corr/abs-1912-02954, DBLP:journals/corr/abs-1901-04620,8406560}. Zhang et al. highlight the importance of game-theoretic soundness by demonstrating wide classes of attacks that affect most of these protocols~\cite{8835227}.

\subsection{Scalability and Network Fairness}
In most existing blockchain literature, it is typically assumed that all the agents have equal access to the network, albeit with some finite delay. However, this is seldom the case in practice, where some nodes may have better internet connections than others as shown by Gencer et al.~\cite{10.1007/978-3-662-58387-6_24}. Jain et al.~\cite{aamas21} show that asymmetry in access to the network can lead to fairness issues among the parties in a Proof-of-Work blockchain. Their paper shows that the deterioration of fairness measures would impact the profitability of the mining operation for some parties. Consequently, the impacted parties would be left with two rational choices: 
\begin{inparaenum}[(i)]
    \item to either quit the mining operation leading to a loss of non-adversarial computational power which reduces the security against Byzantine adversaries, or 
    \item  to adopt strategic behavior to gain more reward, which could not only be a greater threat to security but also lead to a reduction in the performance of the protocol.
\end{inparaenum}
\subsection{Our Goal}
 With this work, our goal is to develop a suitable alternative to existing blockchain protocols that overcomes the aforementioned challenges. Thus, we designed a \emph{scalable} blockchain protocol that is \begin{inparaenum}[(i)]
     \item Secure, 
     \item Game Theoretically Sound, and 
     \item has Network Fairness properties\end{inparaenum}. Table~\ref{comparison} provides a comparison of desirable qualitative features of our protocol with respect to the existing literature.
 
\begin{table}[h]
\centering
\begin{threeparttable}
    \begin{tabular}{cllll}
            \toprule \multirow{2}{*}{Protocol} & \multicolumn{4}{c}{Properties}\\\cmidrule{2-5}
    & Scalable & GT Soundness\tnote{1} & \multicolumn{2}{c}{Network Fairness} \\ & & & $p_f$ Guarantees\tnote{2} & $\alpha_f$ Guarantees\tnote{3}\\\midrule
    Bitcoin & \xmark~\cite{cryptoeprint:2013:881} & \cmark~\cite{cryptoeprint:2018:138} & \xmark~\cite{aamas21} & \xmark~\cite{aamas21} \\ 
   Ethereum & \xmark~\cite{cryptoeprint:2013:881} & \cmark~\cite{cryptoeprint:2018:138} & \xmark~\cite{aamas21} & \xmark~\cite{aamas21}\\ 
   OHIE & \cmark~\cite{ohie} & \xmark~\cite{aamas21} & \xmark~\cite{aamas21} & \xmark~\cite{aamas21}\\ 
\rowcolor{Gray}   \textcolor{blue}{Interlude} & \textcolor{blue}{\cmark} & \textcolor{blue}{\cmark} & \textcolor{blue}{\cmark} & \textcolor{blue}{\cmark}\\\bottomrule
\end{tabular}
\caption{Qualitative Comparison of Interlude with existing blockchain protocols
}
\begin{tablenotes}
    \item [1] Game Theoretic Soundness, which is achieved when following the protocol is the equilibrium strategy for the participants.
    \item [2] Probability of Frontrunning, described formally in Section~\ref{p_f} should be upper-bound by a low value.
    \item [3] Publishing Fairness, described formally in Section~\ref{a_f} should remain close to 1.
    \item [4] \cmark~denotes that the protocol contains the property and \xmark~denotes that the protocol provably \emph{fails} to contain the property.
\end{tablenotes}
\end{threeparttable}
\label{comparison}
\end{table}
\subsection{Our Approach}
 Our blockchain protocol, \proto\ has a relatively simple design yet resolves the above challenges. The key idea is to allow the \emph{parties} to mine multiple \emph{parallel} blocks containing distinct transactions followed by a series block that contains pointers to a set of parallel blocks, synchronizing the entire network (Figure \ref{fig:main}). As the parallel blocks can come in any order, they can be mined at a fast rate without needing to wait for network synchronisation, thus, speeding up the transaction processing. However, there may be forks in parallel blocks that include the same set of transactions. These forks are quickly resolved with a \emph{series} block that takes, on average, more time to mine, providing enough time for the network to get synchronized. Thus, the protocol draws its name from the perspective that between the chaos of parallel blocks, it masquerades an interlude moment of harmony in series blocks that synchronize the network.
 
 Additionally, to prevent parties from slowing down due to forks, we allow them to move their block to a non-conflicting position. Hence, with \proto, we introduce two novel and complementary design ideas:
\begin{itemize}
    \item[\ding{202}] \emph{Sub-chains} - We propose to split transactions into disjoint subsets such that two transactions that belong to different sets can never conflict with each other. Further, we assign these subsets to different sub-chains that are mined parallelly. This idea yields scalability and network fairness in one shot since the number of parties contending for the same transaction set would be split across the subsets.
    \item[\ding{203}] \emph{Umbrella PoW} -  We propose a simple modification to the standard PoW scheme that allows a party to mine a block with PoW that verifies multiple sets of transactions, but the protocol selects only one set in the main chain. The selection is designed in to prevent any strategic deviation by rational parties. This idea allows us to utilize each PoW efficiently since very little PoW is ``wasted'' on orphan blocks because, in the case of forks, a party can publish a block with the same PoW but on a different sub-chain.
\end{itemize}
\subsection{Our Contributions}
In this work, we present the following contributions:
\begin{itemize}
    \item We develop a novel blockchain protocol, \emph{\proto}, that scales to high throughputs without deteriorating security properties.
    \item In order to achieve game-theoretic soundness and network fairness, we pioneer two complementary ideas: \emph{Sub-chains} that divide the transaction set into disjoint subsets and the \emph{Umbrella PoW} scheme.
    \item We prove the resilience of \proto\  against byzantine adversaries that may control up to $1/3^\text{rd}$ of the computational power (Theorem \ref{thm:liveness} and \ref{thm:safety}).
    \item We show that \proto\ is game theoretically sound by using the Rational Protocol Design Framework~\cite{cryptoeprint:2013:496} to show that our protocol is both \emph{strongly attack-payoff secure} and \emph{incentive compatible} (Section~\ref{section:gt}).
    \item We provide network fairness guarantees on Interlude. (Section~\ref{sec:network_fairness})
    \item We also simulate \proto\ to provide real-world performance estimates. \proto\ supports a transaction throughput of more than 1,500 TPS; close to the VISA payment system today.
\end{itemize}

\section{Model and Preliminaries}
We study our protocol in a multiparty setting and make use of the definitions introduced in existing blockchain literature~\cite{cryptoeprint:2014:765}. This section describes our model briefly. 

\subsection{System Model}
We consider a system with $n$ honest parties $\mathcal{P}=\{P_1, P_2, \ldots, P_n\}$ and an adversary $\mathcal{A} = \{A_1, A_2, \ldots, A_m\}$. Each party is identified by its public key - private key pair $\langle pk, sk\rangle$. However, no party is aware if any other party belongs to the adversary or is honest. These parties are interested in maintaining a \emph{Public Transaction Ledger} while deploying \emph{Proof-of-Work} technology.  Additionally, we include $\text{RO}$, i.e.,  a random oracle for simulating the Proof-of-Work.
\subsubsection{Proof-of-Work (PoW)\label{ro}}

In a Proof-of-Work mechanism, the parties are required to obtain a PoW in the form of a nonce that, along with contents of the block, hashes to a value lesser than the predefined target. We assume that the following two worlds are computationally indistinguishable for any Probabilistic Polynomial Time Turing Machine (PPT TM): Real World where an input $X$ is evaluated by a cryptographic hash function $H$ which outputs $H(X)$ and an Ideal World where the same input $X$ is sent to a random oracle $RO$ that follows the assumptions given in Figure~\ref{random_oracle} formally described in the Appendix~\ref{appendix_ro}.

We model the probability of a party obtaining a PoW  as a memoryless or a Poisson process~\cite{cryptoeprint:2013:881,spectre,nakamoto2012bitcoin}.

\subsubsection{Merkle Trees}
Merkle Tree is a tree which uses cryptographic hash functions to allow for efficient integrity check of the data stored in the leaves~\cite{10.1007/3-540-48184-2_32}. One can generate a Merkle proof of the data stored inside a branch by providing the data stored and $\log_2(k)$ off-path hashes where $k$ is the number of nodes from the root to the branch. They can be constructed using the same Random Oracle $\text{RO}$.
\subsubsection{Types of Parties}
\begin{definition}[Honest Party]
A party is said to be \emph{honest} if and only if it strictly follows the protocol.
\end{definition}
\begin{definition}[Rational Party]
A party is said to be \emph{rational} or \emph{self-interested} if it may strategically deviate from the protocol if the deviation is expected to yield a higher reward.
\end{definition}
\begin{definition}[Adversarial Party]
A party is said to be an \emph{adversarial} (and sometimes also known as ``Byzantine'' in the literature) may not follow the protocol. The adversary's goal is to disrupt the operation of the protocol, and it does \textbf{not} try to maximize its reward.
\end{definition}



\subsection{Transaction Ledger}
A blockchain supports transactions that can be programmed to be highly complex as compared to traditional payment systems. For instance, Bitcoin allows multiple inputs multiple outputs ($s$ inputs, $l$ outputs) in a single transaction, which is virtually unheard of in banking systems. Interlude divides these transactions into two types: \emph{typical} and \emph{non-typical} transactions. Simple one-to-many transactions are known as typical transactions while all other types of transactions such as multiple-input transactions, multi-signature transactions and complicated scripts are said to be non-typical transactions. 

\begin{definition}[Transaction]
A transaction $\mathcal{T}(a_1, a_2, \ldots a_s \rightarrow b_1, b_2, \ldots, b_l)$ is associated with public keys $a_1, a_2, \ldots, a_s$ and $b_1, b_2, \ldots, b_l$ transfers a sum the cryptocurrency from the total balance of $a_1, a_2, \ldots, a_s$ to the balance of $b_1, b_2, \ldots, b_l$.
\end{definition}
Non-typical transactions make up only 0.25\% of the transaction corpus in Bitcoin till date~\mbox{\cite{10.3389/fbloc.2019.00007}}. Interlude can replicate all types of transactions supported by Bitcoin and Bitcoin-like cryptocurrencies.
\begin{definition}[Typical Transaction]
A transaction $\mathcal{T}(a \rightarrow b_1, b_2, \ldots b_l)$ is said to be \emph{typical} \emph{iff} it is associated with only one sender.
\end{definition}

\begin{definition}[Public Transaction Ledger]
A \emph{public transaction ledger} $\mathcal{L} = \langle\Pi, R_\kappa(\cdot)\rangle$ is a tuple of protocol $\Pi$ and a transaction acceptance rule $R_\kappa(\cdot)$ such that any honest party $P_i$ will accept a valid transaction $\mathcal{T}$ if $R_\kappa(\mathcal{T}, P_i)$ returns true for a security parameter $\kappa$.
\end{definition}
\subsection{Network Model}
In this work, we consider a \emph{partially synchronous} network model. The security of our protocol depends on the existence of an upper-bound $\delta$ such that any message transmitted by an honest party at time $t$ is received by all other honest parties by time $t + \delta$. Although the protocol is independent of $\delta$ for regular operation, our proofs make use of this bound to prove the protocol's desirable properties. Note that this is the weakest possible assumption since Pass et al.~\cite{10.1007/978-3-319-56614-6_22} show that blockchain consensus is not possible in an asynchronous setting where $\delta$ can be unbounded.
\subsection{Our Goal}
Our primary goal is to scale blockchains and propose a protocol $\Pi_\text{\proto}$ in a fair manner by parallelizing the transaction processing in a blockchain among the miners. The protocol and acceptance rule $R_\kappa(\cdot)$ should form a valid Public Transaction Ledger $\mathcal{L}$. Our secondary goal is to ensure that the blockchain protocol we propose satisfies existing notions of Game-Theoretic Soundness and Network Fairness. We show that there exists a reward scheme $\mathfrak{R}$ under which $\Pi_\text{\proto}$ satisfies Strong Attack-Payoff Security and Incentive Compatibility for Game-Theoretic Soundness, and we determine upper-bound on $p_f$ and approximation of $\alpha_f$ to demonstrate the Network Fairness properties.

In the following section, we start by giving high level overview of our proposed protocol \proto.
\section{Conceptual Design}
\label{sec:cd}

To parrallelize the transaction processing, we propose a design consisting of a sequence of $k$ \emph{parallel} blocks. 
Each of the $k$ parallel blocks contains a disjoint set of transactions. We index the transaction sets via the last bits of the public key that signs a given transaction and refer to these indexes as \emph{sub-chains}.

A parallel block contains the root of the Merkle tree formed by all the transaction sets but only includes a single transaction set and a Merkle proof for the corresponding branch of the Merkle tree. Thus, if one were to modify any transaction in the block, the Merkle root would get invalidated. Additionally, in the case of a fork in parallel blocks, the party can publish a copy of the block on another sub-chain. A double-spending transaction must belong to the same sub-chain since it needs to be signed by the same public key. Since the chain can only contain one block at a particular position in the sub-chain, a valid chain cannot contain any double-spending transactions. Thus, the order in which parallel blocks are mined does not matter; they can be mined without network synchronization.

\begin{definition}[Sub-chain]
A \emph{sub-chain} with index $i$ contains blocks that only include transactions that are signed by a public key that has last $\log_2(k)$ bits as $i$.
\end{definition}

The $k$-set (formally defined in Definition~\ref{kset}) is a set of disjoint sets of transactions and the ability to update a block in case of a fork is what yields our protocol network fairness properties.

The parallel blocks are mined quickly, there could be forks. Due to innovative ideas of sub-chains, any forks at a particular round could only possibly grow exponentially in later rounds. Thus,  we need a new mechanism to resolve these forks. In \proto, once a $k$-set of blocks has been mined, we propose that the parties are required to mine a \emph{series block} containing a pointer to each of these $k$ blocks. A series block has typically higher target and hence takes a long time to  mine. Thus, there are fewer chances of forks in a series block. Thus, we find that the chain with an honest party gets synchronized with the chains available with other honest parties during this time period. In fact, one can intuitively observe that since the probability of a fork in series blocks is independent of the value of $k$, we can safely increase this value to scale the blockchain.

When given a chance to mine a parallel block, the party mines on all the sub-chains at once using the \emph{Umbrella PoW} scheme. Upon mining a block, the party can choose any sub-chain to publish the block and in case of a fork, she can publish the block again on another sub-chain.

\noindent \emph{Umbrella PoW.} Typically, in a PoW scheme, the party is required to obtain a nonce, that along with the hash pointers to previous block(s) and the root of the Merkle tree containing all the transactions in a block hashes to a value less than the given target. We modify this scheme by constructing a Merkle tree having different transaction sets as its branches. The party is now required to not only produce the nonce that hashes to a value less than the target but also a valid Merkle proof to the branch that contains the transaction set which a party wishes to include in the block\footnote{The term Umbrella comes from the fact that a single PoW is sufficient to cover multiple transaction sets.}. 
\begin{figure}
    \centering
    \includegraphics[width=\linewidth]{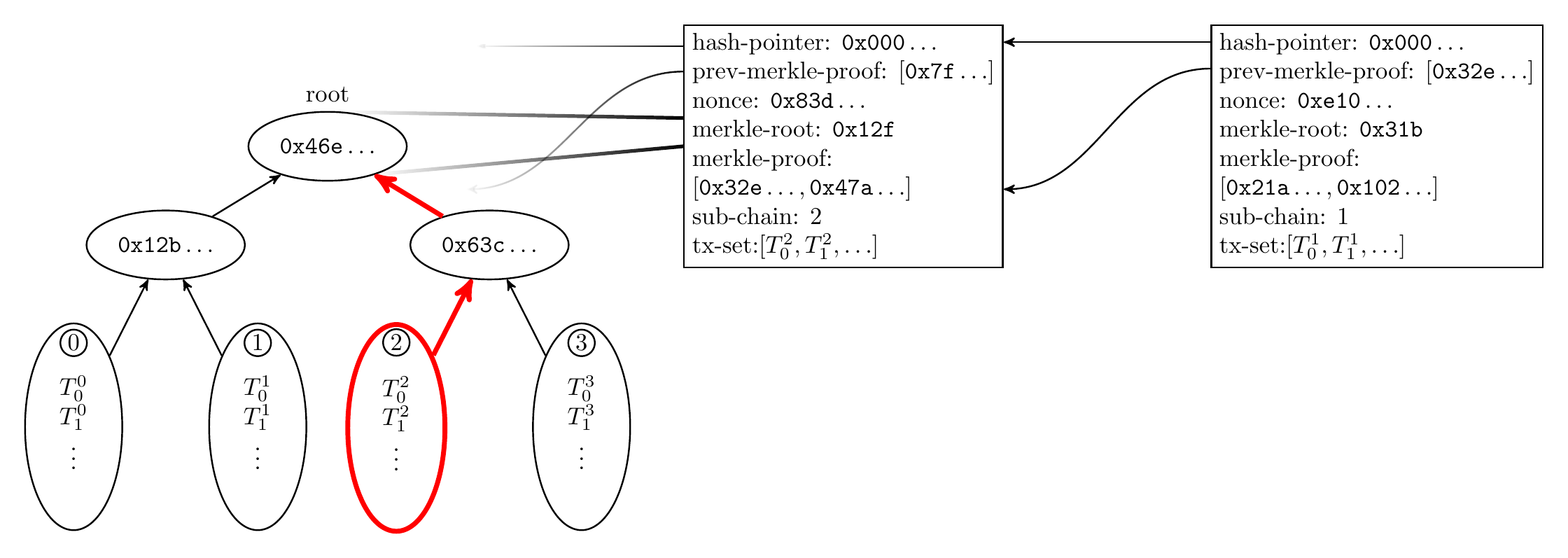}
    \caption{Umbrella PoW Scheme}
    \label{fig:umbrella}
\end{figure}

We now require the next block to contain both, the hash pointer to the previous block and the Merkle proof contained in the previous block to retain the tamper-proof property of PoW. If one were to modify the contents of the block, the Merkle root would get invalidated or if one were to change the branch of the Merkle tree, the nonce of the succeeding block would be invalidated. The Umbrella PoW Scheme is further illustrated in Figure~\ref{fig:umbrella}.

\noindent We now formally describe the construction of our protocol.
\begin{figure}[!h]
    \centering
    \setlength{\fboxsep}{9pt}%
    \fbox{%
    \begin{minipage}{8cm}
    \begin{center}
         \textbf{Protocol $\Pi_\text{\proto}$}
    \end{center}
    \raggedright The protocol $\Pi_\text{\proto}$ is run by parties $\mathcal{P}$ interacting among themselves in a partially synchronous network model \st{for a time $t$} in the presence of the adversaries $\mathcal{A}$, all with the random oracle $\text{RO}$.
    \begin{enumerate}
        \item[\ding{202}] \textbf{Longest Chain Selection} Upon receiving chains $\mathcal{C}_1, \mathcal{C}_2, \ldots, \mathcal{C}_r$ select the chain $\mathcal{C}_\text{max}$ using  Chain Selection Rule (Section \ref{ssec:chain_selection}).
        \item[\ding{203}] \textbf{Mining} If there is a complete $k$-set of parallel blocks at the top of $\mathcal{C}_\text{max}$, the party would mine a series block according to Algorithm~\ref{algo:mining_big} else it would try to add a parallel block to the incomplete $k$-set according to Algorithm~\ref{algo:mining_small}. Whenever a party mines a block $b$, it would update $\mathcal{C}_\text{max}$.
        \item[\ding{204}] \textbf{Updating} If there is a fork detected with one of the parallel blocks mined by a party $P$ and the block mined by party $P$ does not have the lowest hash value, then $P$ will publish another block with same nonce $\eta$ but on a different sub-chain.
        \item[\ding{205}] \textbf{Broadcast} At all times, each party $P$ would broadcast the chain $\mathcal{C}_\text{max}$.
    \end{enumerate}
    \end{minipage}
    }
    \caption{Protocol $\Pi_\text{\proto}$ with parameter $k$}
\end{figure}

\section{Implementation\label{implementation}}
In \proto, there are $k$ parallel blocks followed by a series block\footnote{This terminology is borrowed from electrical circuits.}. We explain how such blocks are created and validated in Section \ref{ssec:blk_cre}. Next we explain how we construct chain from these blocks in Section \ref{ssec:chain_selection}. We describe which transactions from $k$-set are accepted in Section \ref{ssec:trans_acceptance_rule}. Finally, we provide the overview of theoretical guarantees of \proto\ in Section \ref{ssec:proto_guarantees}.

\subsection{Block Creation\label{block_construction}}
\label{ssec:blk_cre}
\subsubsection{Parallel Block}
A parallel block always belongs to a sub-chain. 

\begin{algorithm}
\caption{An algorithm for mining a parallel block}
\label{algo:mining_small}
\begin{algorithmic}
\Procedure{Mine\_ParallelBlock}{top, M, UTXO}
\State $B.\text{top} \gets top$
\Comment{A pointer to the preceding series block.}
\State $B.\text{merkle\_root} \gets M.\text{root}$
\While{\texttt{True}}
    \State {Sample a random nonce $\eta$}
    \If{\Call{$\text{RO}$}{$\eta || B.\text{top} || B.\text{merkle\_root}$} $\leq \text{Target}_\text{parallel}$}
        \State $B.\text{nonce} \gets \eta$
        \State {$B.h \gets$ \Call{$\text{RO}$}{$\eta || B.\text{top} || B.\text{merkle\_root}$}}
        \State {Sample a random $i \in \text{Empty Sub-chains}$}
        \State \Comment{\parbox[t]{.75\linewidth}{Select a random sub-chain among those that don't have any blocks yet}}
        \State $B.\text{sub-chain} \gets i$
        \State $B.\text{merkle\_proof} \gets M.\text{proof}(i)$
        \State \Comment{The Merkle Proof to the $i^\text{th}$ transaction set}
        \State $B.\text{transaction\_set} \gets Tx\_Set(UTXO, i)$
        \State \Return{B}
    \EndIf
\EndWhile
\State \Return{\texttt{null}}
\EndProcedure
\end{algorithmic}
\end{algorithm}

\noindent \emph{Validity.} A valid parallel block is required to contain transactions belonging to only one sub-chain and contain a valid set of transactions along with a single hash pointer to a series block.  
\subsubsection{Series Block}
When given a chance to mine a series block, the party can produce a block containing $k$ hash pointers to parallel blocks that form a $k$-set. The Series Block can contain to leftover transactions from all the $k$ sub-chains as well as more sophisticated types of transactions such as multiple input transactions, scripted transactions, etc.

\begin{algorithm}
\caption{An algorithm for mining a series block}
\label{algo:mining_big}
\begin{algorithmic}
\Procedure{Mine\_SeriesBlock}{top, M, UTXO}
\State $B.\text{top} \gets top$
\State \Comment{\parbox[t]{.9\linewidth}{A list of $k$ hash pointers and Merkle proofs to the $k$-set.}}
\State $B.\text{merkle\_root} \gets M.\text{root}$
\While{\texttt{True}}
    \State {Sample a random nonce $\eta$}
    \If{\Call{$\mathcal{F}_\text{RO}$}{$\eta || B.\text{top} || B.\text{merkle\_root}$} $\leq \text{Target}_\text{series}$}
        \State $B.\text{nonce} \gets \eta$
        \State {$B.h \gets$ \Call{$\text{RO}$}{$\eta || B.\text{top} || B.\text{merkle\_root}$}}
        \State $B.\text{transaction\_set} \gets Tx\_Set(UTXO)$
        \State \Return{B}
    \EndIf
\EndWhile
\State \Return{\texttt{null}}
\EndProcedure
\end{algorithmic}
\end{algorithm}
\noindent \emph{Validity.} A valid series block can contain both typical and non-typical transactions belonging to any sub-chain. They must contain a valid set of transactions that do not conflict with any transaction in any of the preceding blocks. A series block must also contain hash pointers to a valid $k$-set along with their accompanying Merkle proofs from the Umbrella PoW scheme.
\subsection{Chain Construction\label{chain_construction}}
\label{ssec:chain_con}
\begin{figure}[ht]
\centering
\includegraphics[width=0.9\linewidth]{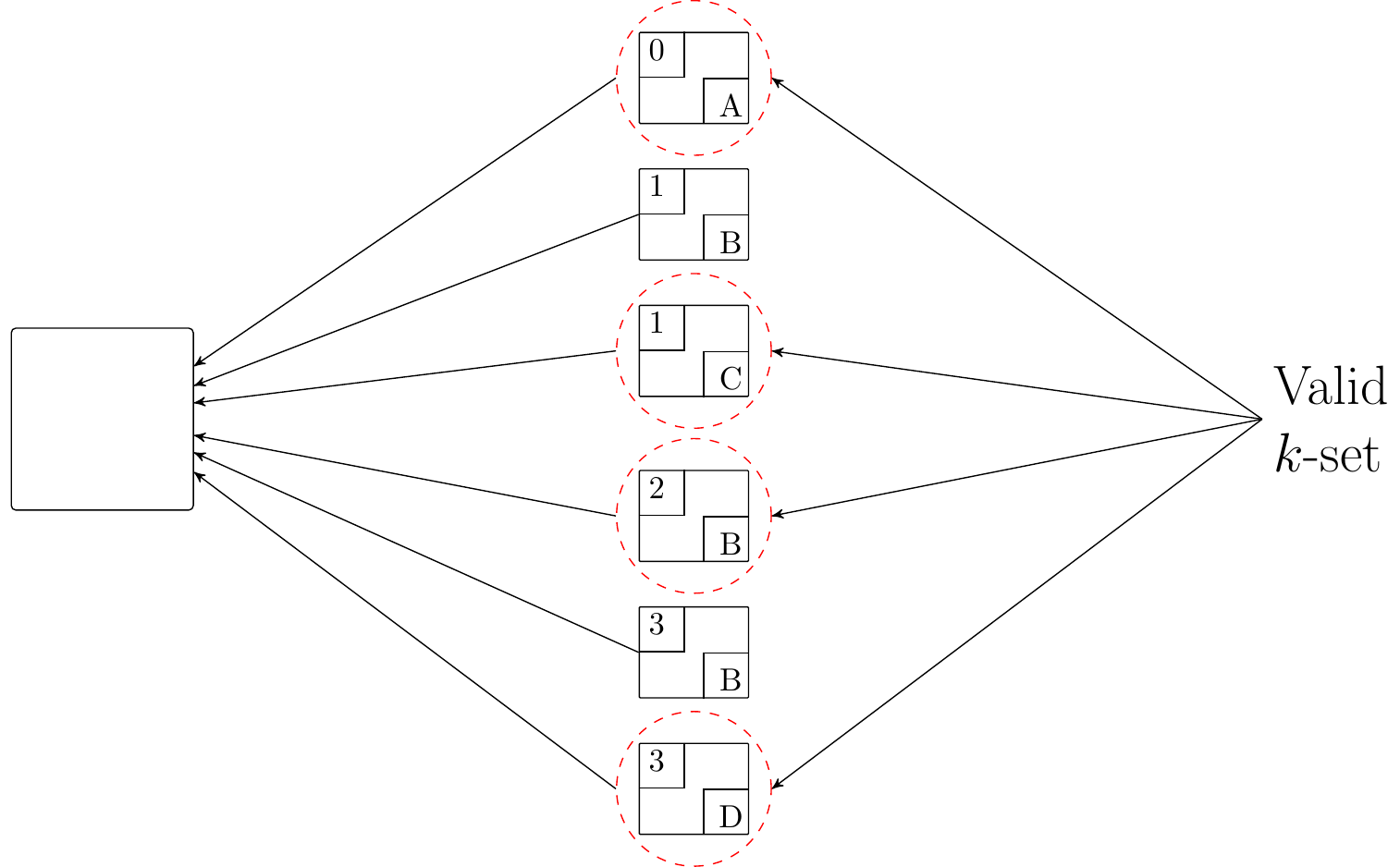}
\caption{Illustration of an example of a round with $k = 4$\label{fig:kset_example}}
\end{figure}

\begin{definition}[$k$-set\label{kset}]
A \emph{$k$-set} is defined as a set of $k$ blocks that share the following properties:
\begin{inparaenum}[(i)]
    \item All blocks have the same $B$.top set, 
    \item no two blocks share the same $B$.chain, and
    \item no two blocks have the same merkle root or nonce.
\end{inparaenum}
\end{definition}
Figure \ref{fig:kset_example} depicts an example of a valid $k$-set. The top-left label denotes the index of the sub-chain and bottom-right label denotes the block such that two blocks with the same bottom-right label are copies of the same block with different transaction set. A valid $k$-set contains only the blocks with distinct sub-chains and merkle-roots.
\subsubsection{Chain Validation}
A chain is considered valid \textit{iff}:
\begin{enumerate}[(i)]
    \item The last series block on the chain contains a hash pointer to a valid $k$-set.
    \item The last series block $B$ contains a valid nonce such that $\text{RO}(\eta || B.\text{top} || B.\text{merkle\_root})\leq \text{Target}_\text{series}$.
    \item The last $k$-set contains a pointer to a series block.
    \item Every parallel block $b$ in the $k$-set contains a valid nonce such that The last series block $B$ contains a valid nonce such that $\text{RO}(\eta || b.\text{top} || b.\text{merkle\_root})\leq \text{Target}_\text{parallel}$
    \item The chain without the last series block and $k$-set is a valid chain.
\end{enumerate}

\subsubsection{Chain Selection\label{ssec:chain_selection}}
The longest chain is selected based on the following parameters (in the order of priority)\footnote{pseudo code description of the rule is in Appendix~\ref{chain_selection_code}}:
\begin{enumerate}[(i)]
    \item The chain with more number of series blocks is preferred.
    \item The chain with more blocks in the incomplete $k$-set is preferred.
    \item In case of two chains having same number of blocks in the incomplete $k$-set, pick the one with the lower hash value for the last series block.
    \item In case of two chains having different $k$-set on top of the same series block, the $k$-set having the parallel block having the least hash value is selected.
\end{enumerate}
The above rules are designed to guarantee that all honest parties unanimously decide to mine on top of a unique chain. In Section \ref{gt_analysis}, we show that picking the prescribed chain is also the Nash Equilibrium strategy. Hence, all rational parties should mine on top of this chain.

The first rule is picks the chain with the maximum PoW. The second rule suggests to pick a chain which has a greater probability of completing the last $k$-set making it more likely to end up as the longest chain. The third rule is a tie breaker rule that unanimously picks one chain. If a party were to deviate while all other parties pick some other serial block, the probability of completing the $k$-set before all other parties would be very low. Hence, a rational party would choose to mine on the chain suggested by these rules.
\subsection{Transaction Acceptance Rule\label{ssec:trans_acceptance_rule}}
We say transactions $ \mathcal{T}_1,\mathcal{T}_2,\ldots, \mathcal{T}_j$, all to be added in round $r$ are valid \textit{if}
$\sum_{i\in[j]} \mid\mathcal{T}_i(\cdot)\mid \leq \sum_{ i \in [s]}\text{Balance}(a_i)$ where $\mid\mathcal{T}_i(\cdot)\mid$ represents the sum of outputs in $\mathcal{T}_i$, and $\text{Balance}(a_i)$ is the amount of unspent currency associated with the public key $a_i$ in all the transactions that precede it in the transaction ordering.

We modify Bitcoin's transaction acceptance rule of accepting any transaction that is $\kappa$ series blocks deep to check if there are any visible forks that may compete with the current longest chain. Hence, the honest party $P_i$ accepts a transaction $\mathcal{T}$ with security parameter $\kappa$ by applying the rule $R_\kappa(\mathcal{T}, P_i)$.\\

\noindent \textbf{$R_\kappa(\mathcal{T}, P_i)$}: 
Let $\mathcal{C}_i$ be the set of all chains available with $P_i$, $C^* = \max_{\{C \in \mathcal{C}_i| \mathcal{T} \in C\}} |C|$, and $C' = \max_{\{C \in \mathcal{C}_i| \mathcal{T} \notin C\}} |C|$, $R_\kappa(\mathcal{T}, P_i)$ returns true \textit{iff} $|C_\text{max}| - |C'| > \kappa$.

\begin{figure}
    \centering
    \includegraphics[width=\linewidth]{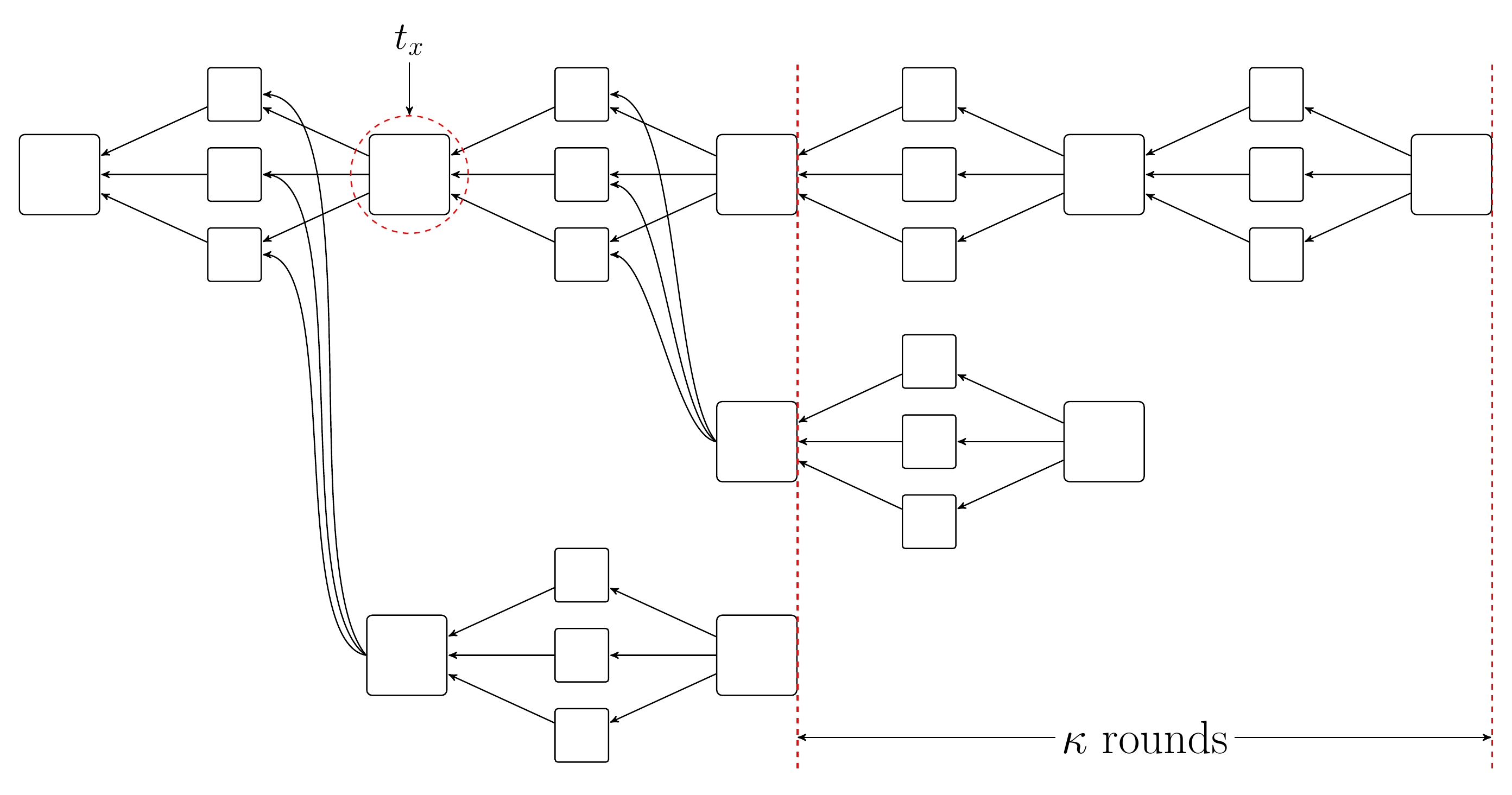}
    \caption{An example of transaction getting accepted with $\kappa = 2$}
    \label{fig:acceptance}
\end{figure}

Figure \ref{fig:acceptance} illustrates an example scenario where a transaction is accepted with $\kappa = 2$. Notice that $\kappa$ is measured only from the longest fork that does not include the transaction.\\
\emph{Transaction Ordering.}
Since each block in \proto\ can only contain valid transactions which do not conflict with any transaction in the preceding blocks and there cannot be two conflicting transactions across parallel blocks on different sub-chains since they must be signed by the same public key. Since a valid chain in \proto\ only contains valid transactions, making it a linear blockchain protocol. In such a case, the transaction ordering is implicit; a transaction precedes another transaction if it is in a block that precedes the block containing the latter transaction.
\subsection{Overview of Guarantees Provided}
\label{ssec:proto_guarantees}
A protocol $\Pi$ implements a robust public transaction ledger if it can ensure the following properties hold for any transaction created by an honest party with a high probability~\cite{cryptoeprint:2014:765}:

\begin{enumerate}
    \label{properties}
    \item \emph{Liveness} with parameters $u, \kappa$: If all parties attempt to include a valid transaction in the ledger at time $t$, then the transaction would be accepted by every honest party with security parameter $\kappa$ at a time $t+u$ with high probability.
    \item \emph{Safety} with parameter $\kappa$: If an honest party has accepted a transaction with a security parameter $\kappa$ at time $t$, then at any time after $t+\delta$ the transaction must be present in the longest chain available with every honest party.
\end{enumerate}
We formally define and prove these properties for \proto\ in Section~\ref{liveness}~and~\ref{safety}.

In Section~\ref{gt_analysis} we prove that $\Pi_\text{Interlude}$ is strongly attack-payoff secure and incentive-compatible for any rational coalition, i.e., if a player is rational, it would choose to follow $\Pi_\text{Interlude}$ over any other action. Hence, we can relax the honest-majority assumption to a rational majority because every rational party would act in the same manner as an honest party. This relaxation is similar to Badertscher et al.'s treatment of the Bitcoin protocol~\cite{cryptoeprint:2018:138}.

In Section~\ref{sec:network_fairness} we formally introduce two measures of network fairness $p_f$ and $\alpha_f$ and show that in \proto\ both these measures are bounded to values that are suitable for fair operation of a distributed ledger.
\section{Security Analysis\label{security_analysis}}
In this section, we show that $\Pi_\text{Interlude}$ implements a secure distributed ledger under the assumption that more than $2/3^\text{rd}$ of the computational power is honest.
\subsubsection*{Overview of Security Analysis}
On a high level, we develop a random walk model to model the adversary's advantage in comparison to the honest parties. First we show that the random walk is biased in favor of the honest majority (Theorem~\ref{thm:liveness}) and then prove the liveness and safety properties (Theorems \ref{thm:liveness} and \ref{thm:safety} respectively). We start by defining the notations that we use in the following text and proving preliminary lemmas followed by theorems that prove the claimed security properties.

A table of notations used in the security analysis is provided in the Appendix~\ref{notation_table}. The complete proofs of lemmas in this section can be found in Appendix~\ref{allproofs}.
\begin{definition}[Round]
A \emph{round} is defined as the time elapsed between mining of a series block by an honest party and the series block that precedes it.
\end{definition}
\subsubsection{Forks}
\begin{figure}[htbp]
\centering
\includegraphics[width=\columnwidth]{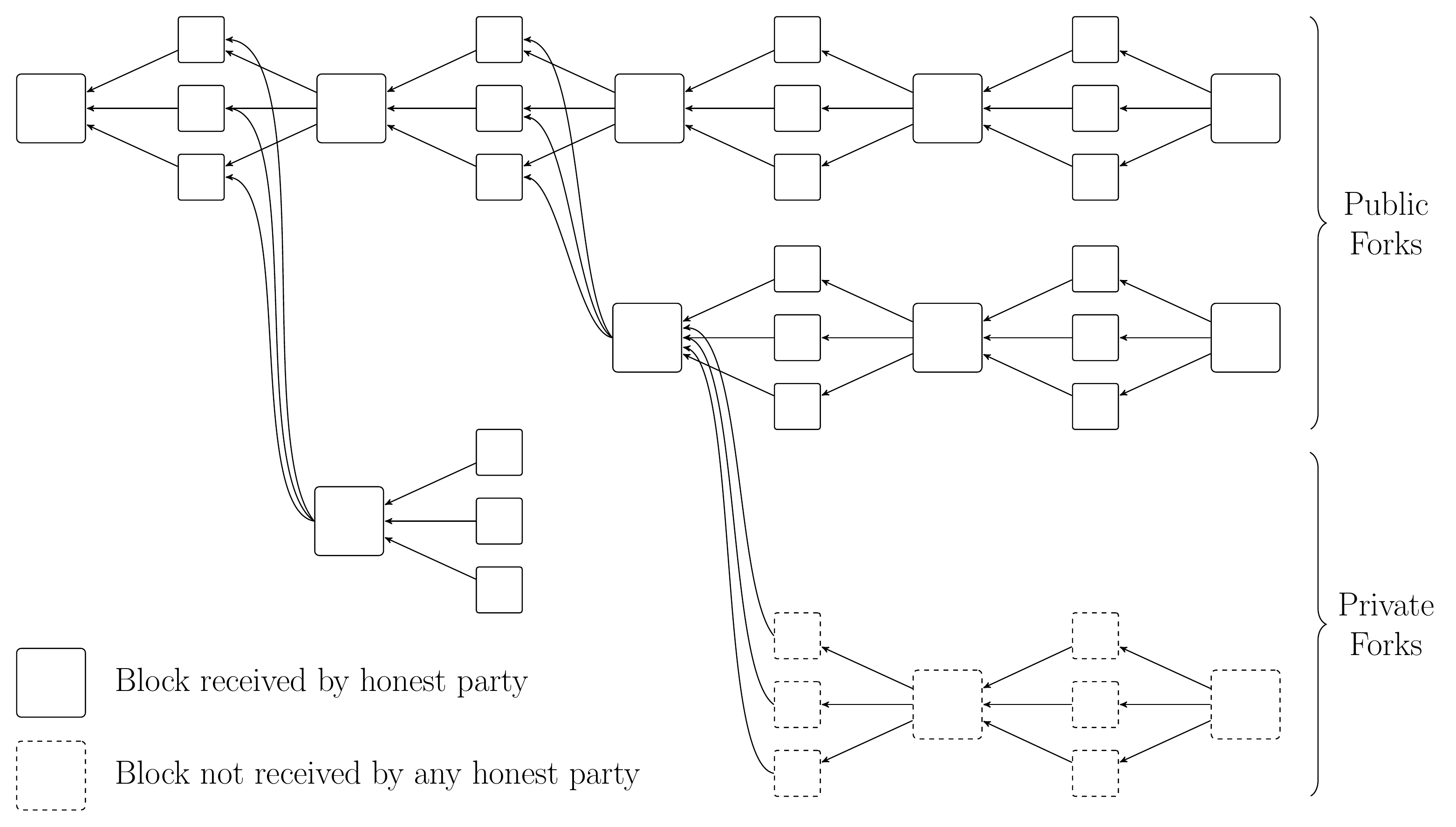}
\caption{Examples of different kinds of forks and their classification.\label{fig:forks}}
\end{figure}

Since we adopt a novel chain selection rule that distinguishes between forks visible to the honest parties and those not visible to the honest parties. We define the two complementary types of forks as follows:
\begin{definition}[Public Fork\label{public_fork}]
A fork that has been adopted by atleast one honest party is said to be a \emph{public fork}.
\end{definition}
On the other hand, a private fork is not adopted by any honest party.
\begin{definition}[Private Fork]
A \emph{private fork} is a fork that is either not completely announced to the honest parties or is lagging behind the longest chain available with every honest party.
\end{definition}
Notice that only a public fork may receive contributions from the honest party while a private fork must only contain contributions from an adversary. Figure \ref{fig:forks} illustrates different kinds of forks and their classification.
\subsection{Background for Security Analysis}
\subsubsection{Assumptions\label{params}}
We assume the following constraints on the system parameters:
\begin{itemize}
    \item The adversary is limited to $1/3$ majority, i.e., $\alpha < 1/2$.
    \item In expectation, the time taken to mine a series block is larger than the time taken to mine $k$ parallel blocks, i.e.,
    $\big(\frac{1}{\beta}-\delta)>\frac{k}{\lambda}$.
    \item The time taken to mine a series block is generously large, i.e., $1/\beta > 20\delta$.
    \item The number of sub-chains is generously large, i.e., $k > 100$.
\end{itemize}

\subsubsection{Important Lemmas}
In this section, we first prove a property of public forks that follows naturally from their definition in Lemma~\ref{eq_fork}. We then show that with each round, the number of forks ($f$) either increases with a probability of at most $1/4+\epsilon$ or remains same or decreases (Lemmas~\ref{moreThanf},\ref{morethanf_1} and~\ref{private_difficulty}).
\begin{lemma}
\label{eq_fork}
All public forks must have the same number of blocks.
\end{lemma}
\begin{lemma}
\label{moreThanf}
If there are $f$ public forks at round $r$, then the probability that by round $r+1$, there are $f$ or more than $f$ forks remaining is less than $1/2+\epsilon$
\end{lemma}
\begin{proof}
The analysis begins by dividing into two possible cases, one in which the honest parties mine exactly one block with probability $1-\epsilon$ and one in which honest parties mine more than one block with probability $\epsilon$. Then we show that in the first case, the probability of advancing more than $f$ forks by one round is upper-bounded by $1/2$. Hence, the total probability is upper-bounded by $1/2+\epsilon$. The complete proof is given in Appendix \ref{proof_appendix}.
\end{proof}
\begin{lemma}
\label{morethanf_1}
If there are $f$ public forks at round $r$, then the probability that by round $r+1$, there are $f+1$ or more than $f$ forks remaining is less than $1/4+\epsilon$
\end{lemma}
\begin{lemma}
    \label{private_difficulty}
    The difficulty of advancing a private fork by one round is greater or same as the difficulty of advancing a public fork.
\end{lemma}
\begin{figure}
    \centering
    \includegraphics[width=\linewidth]{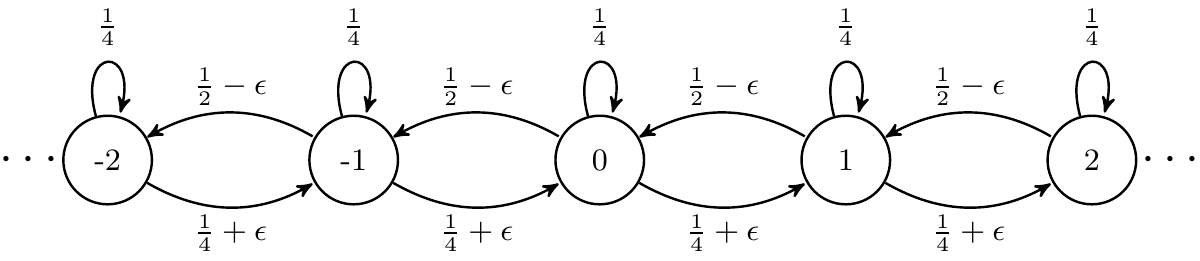}
    \caption{Markov Chain representing the random walk associated with adversary's actions in Brickchain}
    \label{fig:markov_chain}
\end{figure}
\subsection{Liveness\label{liveness}}
We now prove the Liveness property as introduced in Section~\ref{properties}.
\begin{definition}[Liveness]
A ledger $\mathcal{L}$ is said to satisfy \emph{Liveness} with parameters $\kappa, r$ if a \hl{given} transaction $\mathcal{T}$ is included in a block mined by an honest party at round $r_0$, then the transaction would be accepted by every honest party with security parameter $\kappa$ atleast once in rounds $[r_0,r_0+r]$ with a high probability, i.e., $P\big(|C^*| - |C'| > \kappa\big) \geq 1-e^{-\Omega(r-\kappa)}$ at round $r_0+r$.
\end{definition}
\begin{lemma}
\label{lemma:liveness}
In $\Pi_\text{Interlude}$, if for a given transaction $\mathcal{T}$, $C^*$ is one round behind $C'$ at some round $r_0$, then $P\big(|C^*| - |C'| \leq \kappa\big) < e^{-\Omega(r-\kappa)} \forall r\in [r_0,r_0+r]$.
\end{lemma}
\begin{proof}
Let $f \geq 0$ be the number of public forks that do not contain a particular transaction and let $l$ be the number of rounds between the longest fork that does not contain the transaction and the current longest chain adopted by an honest party.
$z = f-l$ denotes the advantage of the adversary over the honest parties. The Markov Chain with $z$ as states is depicted in Figure \ref{fig:markov_chain}. It is a random walk biased towards the left side as $\EX[z]$ decreases with time (from Lemma \ref{moreThanf}, \ref{morethanf_1} and \ref{private_difficulty}). Thus, the honest parties $\mathcal{P}$ have an advantage over $\mathcal{A}$.
The problem of a public fork lagging by $\kappa$ or lesser rounds is equivalent to the above random walk with a cliff at $z = -\kappa$. $P(t,z)$  represents the probability of reaching $z \leq -\kappa$ in the next $t$ rounds. We can solve it as the following recursion.

\begin{equation}
    \label{walk_eqn}\resizebox{0.99\hsize}{!}{$
    P(t,z) = (\frac{1}{2}-\epsilon)\cdot P(t-1,z-1) + \frac{1}{2}\cdot P(t-1,z) 
        + (\frac{1}{4}+\epsilon)\cdot P(t-1,z+1)$}
\end{equation}
with constraints, $\forall \; t$
\begin{equation*}
    P(t, -\kappa) = 1 \quad  \;\; \mbox{and }\;
    P(t, t-\kappa+1) = 0  
\end{equation*}
After solving above analytically, we get the solution as $P(t,z)=Ap^{z-t}$; $p = \frac{\sqrt{16\epsilon^2+12\epsilon+3}-1}{4\epsilon+1}$ and $A = \frac{2-4\epsilon}{3-4\epsilon-2p}$.

Thus, the required probability is $P(r,0) \leq e^{-\Omega(r-\kappa)}$.
\end{proof}
\begin{theorem}
\label{thm:liveness}
$\Pi_\text{Interlude}$ satisfies the liveness property with parameter $\kappa$, $r$ with probability more than $1-e^{-\Omega(r-\kappa)}$.
\end{theorem}
\begin{proof}
When \hl{$C'$} is $\kappa$ or more rounds behind \hl{$C^*$}, an honest party accepts the transaction. 
Note this event being complement of that in Lemma~\ref{lemma:liveness}, the required probability is atleast $1-e^{-\Omega(r-\kappa)}$.
\end{proof}
\subsection{Safety\label{safety}}
We now prove the Safety property of \proto.
\begin{definition}[Safety]
A ledger $\mathcal{L}$ is said to satisfy \emph{Safety} with parameters $\kappa$ if a transaction accepted with security parameter $\kappa$ at a round $r_0$ is present in the longest chain adopted by every honest party in any round $r \geq r_0$ with high probability. Formally, if $|C^*| - |C'| > \kappa$ at round $r_0$ then $P(|C^*| - |C'| \leq 0) \leq e^{-\Omega(\kappa)}$ for any round $r \geq r_0$.
\end{definition}
\begin{lemma}
\label{lemma:safety}
In $\Pi_\text{Interlude}$, \hl{If for a given transaction $\mathcal{T}$, $C'$ is atleast $\kappa$ rounds behind $C^*$ at some round $r_0$, then $P(|C^*| - |C'| \leq 0) \leq e^{-\Omega(\kappa)} \forall: r \geq r_0$.}
\end{lemma}
\begin{proof}
We can observe that $P(|C^*| - |C'| \leq 0)$ can be computing using the random walk in Figure \ref{fig:markov_chain}  with a cliff at $z = \kappa$. We can reuse recursion (Equation \ref{walk_eqn}) for $P(t,z)$, the probability of reaching $z \geq \kappa$ in the next $t$ rounds with the following new constraints: $\forall t$
\begin{equation*}
    P(t, \kappa) = 1  \;\;\mbox{and}\;\;
    P(t, \kappa+1-t) = 0
\end{equation*}
By putting limit $t \to \infty$ and solving the recursion, we get $\lim_{t\to\infty}P(t,0) = p^{-k}$ for $p =\frac{2(1-2\epsilon)}{1+4\epsilon}$.

Thus, $P(|C^*| - |C'| \leq 0) = p^{-k} \in e^{-\Omega(\kappa)}$.
\end{proof}
\begin{theorem}
\label{thm:safety} $\Pi_\text{Interlude}$
 satisfies the safety property with parameter $\kappa$, $r$ with probability more than $1-e^{-\Omega(\kappa)}$.
\end{theorem}
\begin{proof}
For a transaction to be reverted, it is required that there be atleast one fork that does not contain a particular transaction that is longer than the chain adopted by the honest party. This event is negative of the event in Lemma~\ref{lemma:safety} ($|C^*| - |C'| \leq 0$). Thus, a transaction is safe with probability $1-e^{-\Omega(\kappa)}$
\end{proof}

Since $\mathcal{L} = \langle\Pi_\text{\proto}, R_\kappa(\cdot)\rangle$ satisfy both the Liveness and Safety requirements for a distributed ledger, we can say that $\mathcal{L}$ implements a Distributed Ledger.

\section{Game Theoretic Analysis\label{section:gt}}
Recently, researchers have analyzed blockchain protocols from a lens of game theory to highlight the issues faced when the system participants are rational rather than honest. Researchers have discovered various strategies that the parties in a blockchain can adopt to gain more reward than the default strategy. For example, selfish mining is a well-known attack on Bitcoin's incentive mechanism that allows a strategic party to reap more than its fair share of block rewards by waiting to publish its blocks until it causes the most damage to the honest majority~\cite{10.1145/3212998}. Many subsequent papers explored both (i) attacks on Bitcoin’s incentive mechanism~\cite{sapirshtein2016optimal,10.1109/SP.2015.13,10.1145/2976749.2978408,liao2017incentivizing,cryptoeprint:2015:796} and (ii) attacks on other blockchain protocols~\cite{10.1007/978-3-030-53356-4_5, DBLP:journals/corr/abs-1912-02954, DBLP:journals/corr/abs-1901-04620,8406560}. In \cite{8835227}, the authors highlight the lack of systematic game-theoretic analysis of recently proposed blockchain protocols. Often, these attacks not only negatively impact the revenue of other parties but also the security of the blockchain against byzantine adversaries. In fact, the claimed scalability results may not be achievable at the game-theoretic equilibrium. Hence, it is pertinent to develop a reward scheme ensuring the scalability and security claims at equilibrium.

In this section, we analyze the behaviour of a rational party in the system and show that adhering to the protocol is indeed optimal under the appropriate incentives.
\subsection{Overview of Results}
In Subsection~\ref{rewards}, we introduce a reward scheme $\mathfrak{R}$ for \proto. We then use the Rational Protocol Design Framework \cite{cryptoeprint:2013:496} to model the problem of security against a rational adversary as a ``meta-game'' between the protocol designer and the rational adversary. We show that if the protocol designer selects $\Pi_\text{Interlude}$ with reward scheme $\mathfrak{R}$, then a best response of the rational adversary would be to follow the protocol, making the protocol secure at the Nash Equilibrium.
\subsection{Incentives\label{rewards}}
Typically in a PoW blockchain protocol, we have two types of rewards to incentivize  parties to solve the cryptographic puzzles:
\begin{enumerate}
    \item \emph{Block Reward} - The reward parties can assign to themselves for mining a block. This reward mints the underlying currency\footnote{some practitioners may refer it as cryptotokens. For the ease of exposition, we consider it as its equivalent value in fiat currency}, adding to the total amount of currency in circulation. This reward can be fixed suitably to offset the mining costs borne by the party. 
    \item \emph{Transaction Fee} - This is the fee offered by the users for parties' services by providing incentives to include their transactions in the blocks. We denote $\operatorname{Fee}(b)$ as the fees associated with a block $b$.
\end{enumerate}
\noindent \textbf{\proto\ Reward Scheme $\mathfrak{R}$.}
We assume the following constraints on the incentives in the reward scheme:
\begin{itemize}
    \item In \proto, we denote the block reward associated with a series block as $R_{\relbar}$ and that with a parallel block as $R_{\parallel}$.
    \item Typically, the cost of mining $C_\text{mining}$ is majorly contributed by the electricity cost to operate the mining equipment. Hence, one can expect it to be linearly proportional to the time spent in mining a block. Let $\frac{dC_\text{mining}}{dt} = \eta$.
    \item Block rewards exactly cover the expected cost of mining, i.e., $R_{\relbar} = \frac{\eta}{\beta}$ and $R_{\parallel} = \frac{\eta}{\lambda}$.
    \item The transaction fee is much less than the block rewards, i.e., $\operatorname{Fee}(b) < \gamma$ for some constant $\gamma$. This may be enforced by limiting the maximum fees that can be collected by a block.
\end{itemize}
\subsection{Rational Protocol Design Framework} 
In order to argue for the Game-Theoretic soundness of our protocol, we need to show that the best response for a rational adversary or coalition is to follow the protocol. We use the \emph{Rational Protocol Design} (RPD) framework for formally modeling our problem via a Stackelberg game between the protocol designer \texttt{D} and an attacker \texttt{A}~\cite{cryptoeprint:2013:496}. Note that, in this section, attackers are rational agents who may deviate from the protocol to gather more rewards but are not interested in disrupting the protocol otherwise. Typically, the RPD framework models a zero-sum game where the designer's utility is just the opposite of the utility of the attacker. However, we use Badertscher et al.'s modification of the RPD framework for distributed ledgers which uses a non-zero-sum game~\cite{cryptoeprint:2018:138}.
\subsubsection{Attacker's Utility}
The goal of the attacker \texttt{A} is to accumulate the greatest reward possible using a dummy protocol $\phi^{\{\text{RO}\}}$ which provides barebone access to the random oracle RO -- offering the attacker freedom to modify the strategy.

Formally, let the adversary's strategy be $\mathcal{A}$ and the event that block $b$ is inserted in the main chain at round $r$ be denoted by $I^\mathcal{A}_{b,r}$. Then,
\begin{equation}
    U_\texttt{A}(\mathcal{H},\mathcal{A}) = R_\text{block}\cdot Pr(I^\mathcal{A}_{b,r}) - C_\text{mining}
\end{equation}
\subsubsection{Protocol Designer's Utility}
The goal of the protocol designer is to ensure (i) a healthy operation of the distributed ledger, i.e., the safety and liveness properties hold in a given execution, and (ii) the honest parties are rewarded for inserting blocks in the blockchain. Therefore, the utility should incentivize the protocol designer to reward the participants and penalize them for any double-spending attack.

Formally, let the protocol designer's strategy be $\mathcal{H}$ and the event that block $b$ is inserted in the main chain at round $r$ be denoted by $I^\mathcal{H}_{b,r}$ and the event that a double-spending attack occurs be denoted by \texttt{BAD}. Then,
\begin{equation}
    U_\texttt{D}(\mathcal{H},\mathcal{A}) = R_\text{block}\cdot Pr(I^\mathcal{H}_{b,r}) - C_\text{mining}-2^{\text{polylog}(\kappa)}\cdot P(\texttt{BAD})
\end{equation}
\subsubsection{Attack Payoff Security}
The \emph{Attack Payoff Security} property captures that the adversary has no incentive to deviate from the protocol. This notion guarantees that a protocol does not have any incentive driven deviations.
\begin{definition}[Strong Attack Payoff Security (\cite{cryptoeprint:2018:138})]
A protocol $\Pi$ is \emph{strongly attack-payoff secure} if the attacker playing $\Pi$ is $\epsilon$-best response to the protocol designer playing $\Pi$ for a constant $\epsilon$, i.e.,
\[
u_\texttt{A}(\Pi, \phi^{\{\text{RO}\}}) \leq u_\texttt{A}(\Pi, \Pi) + \epsilon
\]
\end{definition}
\subsubsection{Incentive Compatibility}
Not only must an ideal protocol ensure that there are no profitable incentive-driven deviations, but it must also offer an appropriate incentive to the honest parties for following the protocol. The Incentive Compatibility condition captures this.

\begin{definition}[Incentive Compatibility (\cite{cryptoeprint:2018:138})]
A protocol $\Pi$ is said to be \emph{incentive compatible} if $(\Pi, \Pi)$ is a $\epsilon$-Subgame Perfect Nash Equilibrium in the attack game. 
\end{definition}
That is, 
\[
u_\texttt{A}(\Pi, \phi^{\{\text{RO}\}} \mid \mathfrak{H}^r) \leq (1+\epsilon)u_\texttt{A}(\Pi, \Pi\mid \mathfrak{H}^r) 
\forall \; r\] where $\mathfrak{H}^r$ captures the history of the system till round $r$. Intuitively, irrespective of the behaviour till round $r$, it is still an equilibrium to follow the protocol at round $r$.
\subsection{Analysis\label{gt_analysis}}
\begin{lemma}
\label{lemma:greatest_reward}
If the designer plays protocol $\Pi=\Pi_\text{\proto}$ under the reward scheme $\mathfrak{R}$, the attacker's expected utility is upper-bounded by $\gamma$ for any strategy in $\phi^{\{\text{RO}\}}$.
\end{lemma}
\begin{proof}
Let $U^t_\texttt{A}(\Pi,\phi^{\{\text{RO}\}})$ denote \texttt{A}'s utility if the attacker mines a block $b$ at time $t$ and $\mathsf{P} = Pr\big(b\ \text{is accepted}\mid \phi^{\{\text{RO}\}}\big)$.
Consider two cases,\\
\noindent \textbf{Case I:} $b$ is a Series Block
\begin{align*}
    U^t_\texttt{A}(\Pi,\phi^{\{\text{RO}\}}) &= \mathsf{P}\cdot(R_{\relbar} + \operatorname{Fee}(b)) - \eta t\\
    \EX_t[U^t_\texttt{A}(\Pi,\phi^{\{\text{RO}\}})] &= \mathsf{P}\cdot\operatorname{Fee}(b) \leq \gamma
\end{align*}\\
\noindent \textbf{Case II:} $b$ is a Parallel Block
\begin{align*}
    U^t_\texttt{A}(\Pi,\phi^{\{\text{RO}\}}) &= \mathsf{P}\cdot(R_{\parallel} + \operatorname{Fee}(b)) - \eta t\\
    \EX_t[U^t_\texttt{A}(\Pi,\phi^{\{\text{RO}\}})] &= \mathsf{P}\cdot\operatorname{Fee}(b) \leq \gamma
\end{align*}
\end{proof}
\begin{lemma}
\label{lemma:lowest_reward}
If the designer plays protocol $\Pi=\Pi_\text{\proto}$ under the reward scheme $\mathfrak{R}$, and the attacker cooperates by also playing $\Pi_\text{\proto}$ then, the attacker's expected utility is lower-bounded by $0$.
\end{lemma}
\begin{proof}
Let $U^t_\texttt{A}(\Pi,\Pi)$ denote the utility if the attacker mines a block at time $t$. Since the attacker is extending the longest chain, $\mathsf{P} = P\big(b\ \text{is accepted}\mid \Pi_{\proto}\big) = 1$,\\
\noindent \textbf{Case I:} $b$ is a Series Block
\begin{align*}
    U^t_\texttt{A}(\Pi,\Pi) &= \mathsf{P}\cdot(R_{\relbar} + \operatorname{Fee}(b)) - \eta t\\
    \EX_t[U^t_\texttt{A}(\Pi,\Pi)] &= \mathsf{P}\cdot\operatorname{Fee}(b) = \operatorname{Fee}(b) \geq 0
\end{align*}\\
\noindent \textbf{Case II:} $b$ is a Parallel Block
\begin{align*}
    U^t_\texttt{A}(\Pi,\Pi) &= \mathsf{P}\cdot(R_{\parallel} + \operatorname{Fee}(b)) - \eta t\\
    \EX_t[U^t_\texttt{A}(\Pi,\Pi)] &= \mathsf{P}\cdot\operatorname{Fee}(b) = \operatorname{Fee}(b) \geq 0
\end{align*}
\end{proof}
\begin{theorem}\label{saps}
$\Pi_\text{\proto}$ is strongly attack-payoff secure under the reward scheme $\mathfrak{R}$.
\end{theorem}
\begin{proof}
From Lemmas \ref{lemma:greatest_reward} and \ref{lemma:lowest_reward}, 
$\Pi_\text{\proto}$ is strongly attack-payoff with $\epsilon=\gamma$. 
\end{proof}
\begin{theorem}
$\Pi_\text{\proto}$ is Incentive Compatible under $\mathfrak{R}$ if $\operatorname{Fee}(b) \geq \rho$ such that $\rho \geq 2^{\text{polylog}(\kappa)}\cdot P(\texttt{BAD}\mid \Pi_\text{\proto})$.
\end{theorem}
\begin{proof}
The proof follows on similar lines to Lemma \ref{lemma:greatest_reward} and \ref{lemma:lowest_reward} by computing $\EX_t[U^t_\texttt{D}(\Pi_\text{\proto},\Pi_\text{\proto})]$ and showing that it is $\in [0,\gamma]$. Thus, $(\Pi_\text{\proto},\Pi_\text{\proto})$ is a $\gamma$-Subgame Perfect Nash Equilibrium in the attack game as at time $t$ these computations are independent of the history.
\end{proof}

From Theorem~\ref{thm:safety}, $P(\texttt{BAD}\mid \Pi_\text{\proto}) \leq e^{-\Omega(\kappa)}$. Hence, we can always find always find an appropriate $\kappa$ such that $\rho \leq \gamma$. Hence, there will always exist a suitable reward scheme $\mathfrak{R}$ for any configuration of $\Pi_\text{\proto}$.
\section{Network Fairness\label{sec:network_fairness}}
In a blockchain protocol, each node must be incentivized to act honestly. To ensure this not only the network of nodes as a whole is provided enough incentives, but each node in the network should also be provided the correct incentive. Hence, the blockchain protocol must be fair to the participants. Researchers have shown that many widely used blockchain protocols are not fair to parties. 

The authors of~\cite{aamas21} show that asymmetry in network access could result in a lack of fairness to the parties, which may be further exacerbated when the blockchain protocol is scaled to higher throughputs. Using a game-theoretic analysis, they also show that most existing blockchain protocols fail to address this issue. For instance, they show that the OHIE protocol fails to make any progress in the presence of rational parties because of their strategies.

\noindent \emph{Party Contention.\label{miner_contention}}
In a typical scalable blockchain protocol, the block creation rate can be increased without impacting the security against a byzantine adversary. However, this implies that blocks are being mined faster than they are propagated through the network. Since all parties are trying to include the same set of transactions, it is highly probable that two parties would mine blocks containing the same transaction. Although both their blocks may become part of the main chain, only one of them would receive the corresponding transaction fees. Hence, the party that hears a transaction before others receive an unfair advantage, and the party that can broadcast its block faster than others receive an unfair advantage. We identify this as the root cause of network fairness issues in a blockchain protocol and propose to solve the same by dividing the transaction set into disjoint subsets

This issue is evident in an example flaw pointed out in~\cite{aamas21}. OHIE~\cite{ohie} is a blockchain protocol that consists of multiple chains that are synchronized via the hash pointers chosen by the party. Since parties are contending for the same transaction set, a rational party would prefer to place the block as early in the chain as possible to collect the transaction fees as opposed to place it as late as possible prescribed in the protocol. As the authors of~\cite{aamas21} show if the parties were to act rationally, not only would OHIE perform poorly, but the security might also be negatively impacted. Similar game-theoretic flaws have been pointed out in other blockchain protocols in~\cite{8835227}. Hence, maintaining network fairness to prevent such issues is essential for any blockchain protocol.
\subsubsection{Probability of Frontrunning\label{p_f}}
\begin{definition}[$p_f$ (due to~\cite{aamas21})]
\emph{$\{p_f\}^{M}_{m}$} is defined as the probability that some node in the top $M$ percentile (in terms of network speed) manages to create a block that includes a transaction before all nodes in the bottom $1-m$ percentile receive the transaction.
\end{definition}
 If $p_f$ is high, the faster nodes would consistently be able to grab high-value transactions while the slower ones would only be able to pick low-value ones left out by others. Thus, a high $p_f$ would negatively impact some parties' revenue. Since $p_f$ is a measure of unfairness, the lower the value the better it is for the participants in the system. For most other blockchain protocols, as the block creation rate is increased, $p_f \to 1$.

\begin{theorem}
$\{p_f\}^{M}_{m} \leq \frac{M\beta d}{1-\beta\delta}$ for \proto\ where $d$ is the delay between the top $M$ and bottom $m$ percentiles to receive a transaction.
\end{theorem}
\begin{proof}
Let us consider a party in bottom $m$ percentile mining a parallel block at time $t$, then the probability that a party in the top $M$ percentile has mined a parallel block that belongs to the same chain in this time period is $1-e^{-\frac{M\lambda d}{k}} \leq \frac{M\lambda d}{k}$. Therefore,
\begin{equation*}
    \{p_f\}^{M}_{m} \leq \frac{M\lambda d}{k} \leq \frac{M\beta d}{1-\beta\delta}
\end{equation*}

The same probability for a series block is upper-bounded by $\epsilon \leq \frac{M\beta d}{1-\beta\delta}$. Thus, for both the blocks, $\{p_f\}^{M}_{m}$ is guaranteed to remain quite low.
\end{proof}
For most practical cases, $d < \delta$ and our assumption on security parameters requires $\beta\delta \leq 1/20$. Hence, $\{p_f\}^{M}_{m}$ is less than $1/19$ at any block creation rate for any $M$ and $m$. Thus, we can claim that \proto\ guarantees Network Fairness against frontrunning of transactions.
\subsubsection{Publishing Fairness\label{a_f}}
Publishing Fairness ($\alpha_f$) quantifies the advantage a node might have over others in broadcasting a block. If a node is able to propagate its block faster than others, in case of an eventual fork, its block would have a higher probability of being accepted~\cite{aamas21}.
\begin{equation}
    \label{eqn:publishing_fairness}
    \alpha_f(A,B) = \frac{P(\frac{\text{A's block getting accepted}}{\text{A and B mine a block simultaneously}})}{P(\frac{\text{B's block getting accepted}}{\text{A and B mine a block simultaneously}})}
\end{equation}
The issue of publishing fairness is similar to the frontrunning. In this case, two parties contend to publish a block in the same position. Notice that in \proto, in case of a fork in parallel blocks, a party can almost always get their block accepted by shifting it to another sub-chain. Secondly, for series blocks, the probability of mining two blocks consecutively in time $\delta$ is negligible ($\epsilon$). The same is consistent with our experiments since we find that more than 95\% of the blocks that were mined, ended up in the final chain which is close to Ethereum today. Hence, our protocol has an acceptable value of $\alpha_f$.
\section{Evaluation}
In order to empirically verify our claims, we implement a prototype of Interlude in C++ and run the same by simulating a network delay of 40 seconds\footnote{Source code can be shared on request.}. Note that the delay used in our experiments is much higher than that of the Bitcoin network today, which is able to propagate blocks to 95\% of the nodes in less than 10 seconds. In this section, we discuss the operational characteristics of the protocol.

\begin{figure*}[ht]
    \centering
    \minipage{0.32\linewidth}
        \includegraphics[width=0.9\linewidth]{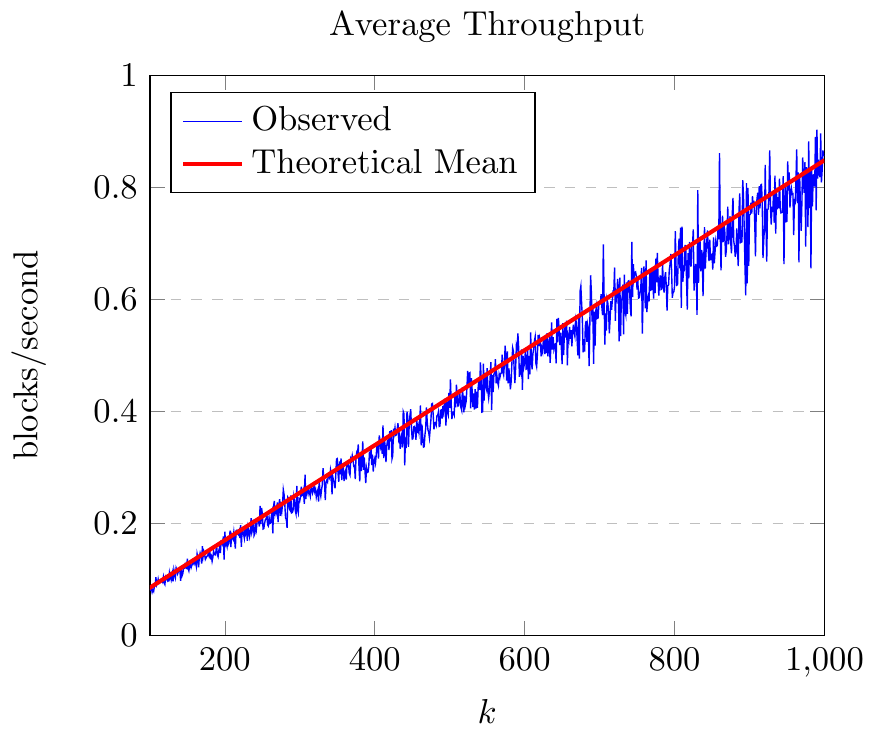}
        \caption{Throughput\label{fig:throughput} as we increase $k$ with $\delta = 40$ seconds, $\lambda/k = 1/\beta - \delta$}
    \endminipage\hfill
    \minipage{0.32\linewidth}
        \includegraphics[width=0.9\linewidth]{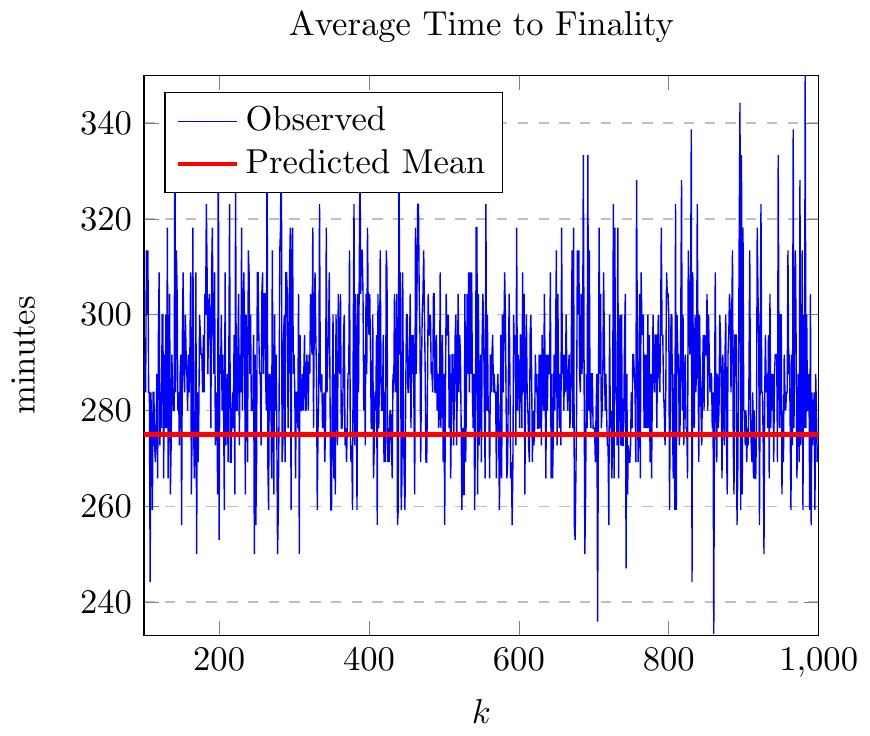}
        \caption{Confirmation Delay\label{fig:confirmation} as we increase $k$ for $\kappa = 14$ and $\delta = 40$ seconds}
    \endminipage\hfill
    \minipage{0.32\linewidth}
        \includegraphics[width=0.9\linewidth]{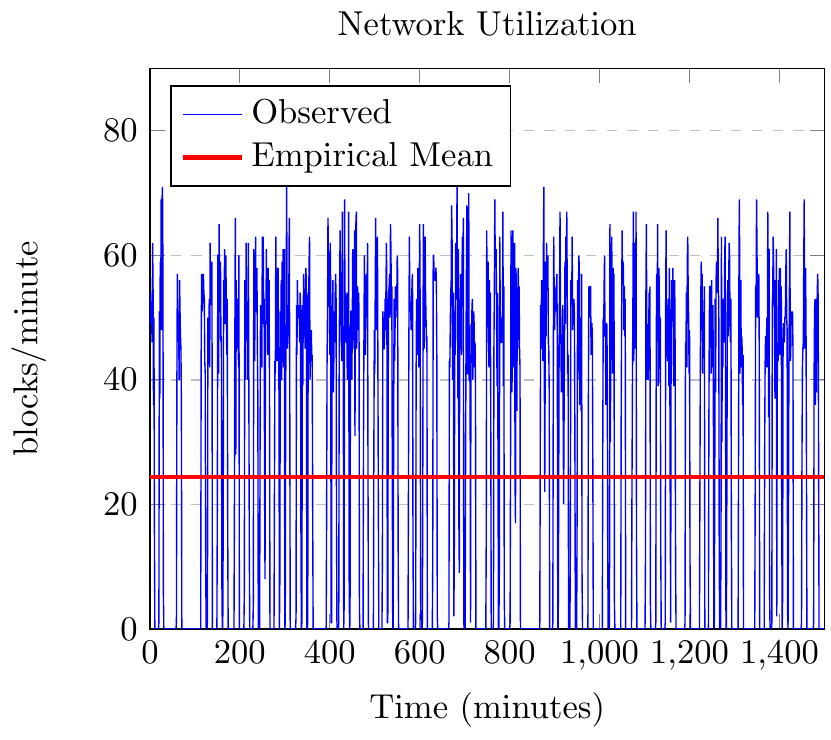}
        \caption{Network utilization\label{fig:network} for $k = 500$}
    \endminipage\hfill
\end{figure*}

\subsection{Transaction Rate}
\begin{table}[h]
\centering
    \begin{tabular}{clll}
    \toprule
    \multirow{2}{*}{Protocol} & \multicolumn{3}{c}{Performance Parameters}\\\cmidrule{2-4}
     & $\text{TPS}_\text{BW}$ & $\text{BW}$ & $\eta = \frac{\text{TPS}_\text{BW}}{\text{BW}}$\\\midrule
   Bitcoin & 7 tx/s & 0.5 MB/s & 14 tps/MBps \\ 
  Ethereum & 10 tx/s & 0.5 MB/s & 20 tps/MBps \\ 
  OHIE & 2,420 tx/s & 2.5 MB/s & 968 tps/MBps \\ 
  \rowcolor{Gray}\textcolor{blue}{Interlude} & \textcolor{blue}{1,500 tx/s} & \textcolor{blue}{2 MB/s} & \textcolor{blue}{750 tps/MBps} \\ 
 VISA Network & 1,700 tx/s & N/A & N/A \\ 
\end{tabular}
\caption{Quantitative Comparison of Interlude with existing blockchain protocols}
\label{comparison_quant}
\end{table}
Arguably the most important metric for measuring real-world performance of a blockchain is the number of transactions per second (tps) it offers.
In our protocol, since a transaction can be included at most once in a valid chain, and each block can consist of a constant number of transactions, we use the average block creation rate as the proxy for measuring transaction rate.

\noindent\emph{Results.} The best case expected throughput for our protocol is $\frac{\beta(k+1)}{2-\beta\cdot\delta}$ in case there are no forks. We achieve comparable block creation rate in our experiments. Figure \ref{fig:throughput} provides evidence that \proto's transaction rate scales linearly with the number of sub-chains employed. Typically, the block size is 1 MB for Bitcoin, which can accommodate nearly 1,500 transactions. Thus, with nearly 1 block per second block creation rate, we claim to rival VISA network's 1,700 transactions per second performance.
\subsection{Time to Finality}
\emph{Time to finality} refers to the time taken for a transaction to be confirmed with security parameter $\kappa$. We observe that the round duration remains constant (approximately $\frac{2}{\beta}-\delta$), hence most transactions should be accepted with security parameter $\kappa$ in around $\kappa \cdot (\frac{2}{\beta}-\delta)$ time. Note that the delay is independent of $k$, which is further verified by our simulations (Figure \ref{fig:confirmation}).

\noindent\emph{Results.} We experimentally use $1/\beta = 600$ which is close to Bitcoin's difficulty and find out that our protocol requires around 1.8x the time required by Bitcoin to confirm a transaction with the same probability for any value of $k$. Figure \ref{fig:confirmation} plots the time to finality vs. $k$, averaged across duration of simulation for $\kappa = 14$ (required to ensure that less than 1 in 1000 chance of reverting the transaction if adversarial computing power 25\%). For comparison, Bitcoin requires 150 minutes to provide the same guarantee.

\subsection{Network Utilization}
So far, we have assumed that the parameter $k$ can be unbounded. However, in practice, increasing $k$ beyond a certain threshold may overwhelm the network capacity and cause a non-trivial increase in $\delta$. If we plot the network utilization with time, we obtain a periodic plot as shown in Figure \ref{fig:network}. We can observe that around 50\% of the time there is a twice the average network utilization and around 50\% of the time the network is nearly idle. This matches the expectation that there would be high network utilization while mining $k$ parallel blocks while the network would be relatively silent while mining a series block.

 Further, experiments in~\cite{ohie} show that the impact of parallel transmission is minimal on block delay unless the network bandwidth is saturated to a high threshold, validating our network simulations.

\noindent \emph{Transaction Propagation.}
In a typical cryptocurrency, the same network is used for  broadcasting both, blocks and  pending transactions. Hence, each party would receive a transaction twice, once before it is included in a block and once inside a block. Thus, the cryptocurrency's actual network utilization would be twice that of the blockchain protocol. So practically, a protocol cannot  use more than half the available network bandwidth.

If $M$ is the maximum network bandwidth, then \proto\ manages to use $M/2$ bandwidth on average. The other $M/2$ can be used for broadcasting pending transactions.

\subsection{Performance Comparison}
We compare the performance of Interlude against other blockchain protocols in Table~\ref{comparison_quant}\footnote{The values in this table are taken from their respective publications or publicly available blockchain statistics.}. We consider three parameters in our comparison, the claimed throughput $\text{TPS}_\text{BW}$ which depends upon the bandwidth $\text{BW}$ at which the protocol was simulated. Since different protocols were evaluated by their respective authors at different bandwidths, we define a \emph{bandwidth efficiency index} $\eta = \frac{\text{TPS}_\text{BW}}{\text{BW}}$ to compare the throughput offered by the protocols at the same bandwidth. For scalable protocols, throughput would increase proportionally with bandwidth with rate $\eta$.

\subsection{Safety}
\begin{figure}
    \centering
    \includegraphics[width=0.825\linewidth]{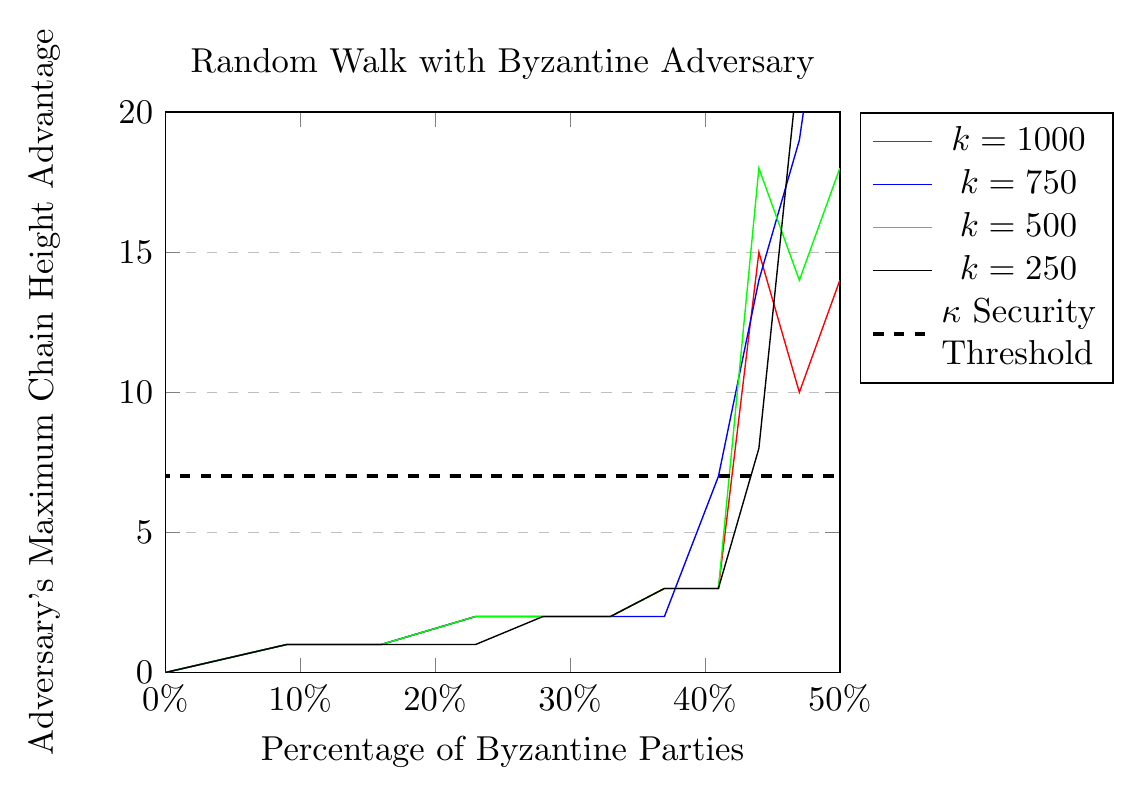}
    \caption{Maximum safety parameter required as we increase the fraction of byzantine parties}
    \label{fig:byzantine_walk}
\end{figure}
So far, we have discussed the performance of \proto\ in a setting with only honest parties. In order to evaluate the safety against byzantine adversaries, we simulate an adversarial strategy similar to~\cite{sapirshtein2016optimal} at varying computing power with the adversary. We describe this strategy in detail in Appendix~\ref{adversarial_strategy}. 

\noindent\emph{Results.} We consistently observe that for any $k$ our protocol is safe upto $33\%$ of byzantine computing power. This further demonstrates the validity of claims presented in Section~\ref{security_analysis}.
\section{Related Work\label{related_work}}
Ever since the first blockchain protocol, Bitcoin, was proposed in 2008, the timeless design of Nakomoto consensus has been thoroughly studied by researchers to identify certain limitations~\cite{nakamoto2012bitcoin,cryptoeprint:2013:881}. Many researchers have also devised new blockchain protocols to overcome scalability, security, and centralization issues.
\subsection{Limitations of Nakomoto Consensus}
Blockchain protocols based on the original Nakomoto consensus, such as Bitcoin and Ethereum, operate at relatively lower throughputs than their centralized counterparts, e.g., VISA Network. Researchers have discovered two major challenges in scaling these protocols:

\subsubsection{Security-Scalability Tradeoff}
Garay et al. \cite{cryptoeprint:2014:765} and Kiayias et al. \cite{cryptoeprint:2015:1019} show that Nakomoto consensus suffers from a loss of security properties as we scale the system.

\subsubsection{Network Fairness}
In practice, all parties participating in a blockchain protocol may not have equal access to the network. \cite{10.1007/978-3-662-58387-6_24} show that there is an inequality in the parties' network access which may affect the system's robustness against attacks. \cite{aamas21} shows that a lack of network fairness can not only negatively impact the profitability of a party but can also lead to strategic deviations, which can be harmful to the security of a blockchain. We use their fairness measures to show that our protocol guarantees network fairness even as we scale to greater throughputs. They also show that this issue plagues most existing blockchain protocols. For instance, they show that~\cite{ohie}, a recently proposed blockchain protocol, fails to make progress under real-world network conditions. This is the primary issue we wish to address with Interlude since it has not been addressed by any other blockchain protocol to date. We provide fairness guarantees on \proto~in Section~\ref{sec:network_fairness}.

\subsection{Scalable Blockchain Protocols}
In this subsection, we discuss the existing blockchain protocols and their design features. Blockchain protocol can be broadly classified into two types: (i) Linear blockchain protocols and (ii) Non-linear blockchain protocols. Linear blockchain protocols are structurally similar to Nakomoto consensus that resembles a linear chain of blocks that grows with time, while Non-linear blockchain protocols adopt a DAG-based architecture (Directed Acyclic Graph). The most distinctive design feature between them is the ability to include conflicting blocks in the main chain. While non-linear blockchain protocols can include conflicting blocks in the main chain, a linear blockchain protocol only includes non-conflicting blocks in the main chain by orphaning the rest of the blocks. Thus, non-linear blockchain protocols require an explicit block ordering algorithm while this order is implicit in linear blockchains. Notice that linear blockchains are a sub-class of non-linear blockchain protocols.

\subsubsection{Linear Blockchain Protocols}
Bitcoin-NG~\cite{eyal2016bitcoin} and GHOST~\cite{cryptoeprint:2013:881} are some well-known examples of linear blockchain protocols apart from the original Nakomoto consensus. Interlude also falls in this category. Thus, our design retains this feature from the original Bitcoin design. Linear blockchain protocols are typically difficult to scale due to formation of forks at high block creation rates or large block sizes. Interlude addresses this issue by increasing the block creation rate via parallel blocks but allow the network to synchronize and resolve forks by shifting blocks across forks and having series blocks that give the network time to synchronize.

\subsubsection{Non-linear Blockchain Protocols}
\cite{eyal2016bitcoin,cryptoeprint:2013:881,spectre,phantom,li2020decentralized,popov2018tangle} are examples of some non-linear blockchain protocols. Chen et al.~\cite{chen2021nonlinear} use game theory to establish the tradeoff between full verification, scalability, and finality-duration in non-linear blockchains. Their result relies on the computational complexity of verifying a blockchain, while in most distributed systems, the message complexity is considered a more important factor. Hence, their bounds are more than optimistic for most practical applications of blockchain protocol. Their result is interesting since it describes the limit of what these blockchain protocols can achieve. These blockchain protocols can be further categorized into subclasses based on their design features (non-exhaustive).
    \\\emph{Combining Multiple Chains.} One of the techniques to scale blockchains is to use multiple instances of Nakomoto consensus that parties mine in parallel while linking the chains in a manner that allows PoW in one chain also to verify blocks in another chain. Thus, by sharing PoW across the instances, we can distribute mining across the chains without dividing the security. OHIE~\cite{ohie} and Chainweb~\cite{Martino2018ChainwebA} use this approach.
    \\\label{sharding}\emph{Sharding.} Another common technique is to use the ``divide-and-conquer'' paradigm. The parties are distributed across shards with some intersection between the shards. These protocols are typically more complex and do not achieve full verification. Since a transaction is only verified by a subset of parties in a shard, these protocols do not achieve full verification. \proto~on the other hand, achieves full verification. Ethereum 2.0~\cite{ethereum.org_2021}, OmniLedger~\cite{8418625}, RapidChain~\cite{10.1145/3243734.3243853}, and PolyShard~\cite{9141331} are examples of such protocols.
\label{nonlinear_issues}
\hl{The non-linear or DAG-based approach suffers from the following drawbacks:}
\begin{enumerate}
    \item Vulnerable to \emph{Balance Attacks} wherein an adversary may partition the network into balanced subgraphs and start mining on one subgraph to invalidate another that contains an accepted transaction~\cite{DBLP:journals/corr/abs-2012-06128}. Interlude is immune to this attack since a party can chose to mine on only one chain as compared to a non-linear blockchain allowing the party to mine on top of \emph{all} the chains at once.
    \item Few blockchain protocols such as SPECTRE~\cite{spectre} and IOTA~\cite{popov2018tangle} can only guarantee pairwise ordering of transactions which may lead to Condorcet cycles in transaction ordering.
\end{enumerate}
We further refer the reader to a comprehensive survey of blockchain-based consensus protocols by Bano et al.~\cite{10.1145/3318041.3355458}.

\subsection{Game Theory in Blockchains}
Blockchain is a distributed ledger that uses mechanism design to ensure correct functioning. Unlike the Visa Network or centralized banking systems, a blockchain is maintained by a decentralized network of nodes. Instead of using a central authority, it incentivizes participants to run and secure the underlying consensus protocol. So, it is crucial to ensure that the nodes that maintain the ledger are appropriately incentivized in order to prevent them from behaving maliciously. Liu et al.~\cite{liu2019survey} provide an overview of the usage of game theory in blockchains. Badertscher et al.~\cite{cryptoeprint:2018:138} propose modifications of the Rational Protocol Design Framework to analyse reward schemes in blockchain protocols that we use to analyse \proto\ in Section~\ref{section:gt}.
\subsubsection*{Incentive-Driven Deviations}
There has been a substantial amount of literature devoted to discovery of strategic deviations in blockchain protocols. Selfish mining family of attacks allow the attacker to grab greater fraction of the reward~\cite{10.1145/3212998,sapirshtein2016optimal,10.1109/SP.2015.13,gobel2016bitcoin,7467362}. Undercutting attacks allow a party to grab greater reward in the absence of a block reward at the expense of security of the blockchain~\cite{10.1145/2976749.2978408,liao2017incentivizing,cryptoeprint:2015:796}. Few others have described various attacks for different blockchain protocols~\cite{10.1007/978-3-030-53356-4_5, DBLP:journals/corr/abs-1912-02954, DBLP:journals/corr/abs-1901-04620,8406560}. Zhang et al. highlight the lack of systematic game-theoretic analysis of recently proposed blockchain protocols that leaves them vulnerable to a broad category of incentive-driven attacks~\cite{8835227}. Towards this, we include a formal game-theoretic analysis of Interlude using the the Rational Protocol Design (RPD) framework and show that it is resistant to incentive driven deviations.

\section{Conclusion}
In this paper, we proposed \proto, a blockchain protocol with an efficacious yet straightforward design. We captured party contention as the root cause of issues that arise due to a lack of network fairness and pioneer two complementary ideas, sub-chains that divide the transaction set into disjoint subsets and the Umbrella PoW scheme, to tackle these issues. We formally proved the security properties of \proto\ to show that it can be used to construct a scalable distributed ledger. We also define a reward structure and use formal game-theoretic analysis to show that it deters incentive-driven deviations. We additionally prove that underlying network conditions would not impact the fairness of our protocol. Finally, we implement our protocol on a network simulator to
\begin{inparaenum}[(i)]
    \item study its operational characteristics,
    \item to empirically validate our theoretical claims on scalability and
    \item test the security against byzantine adversaries.
\end{inparaenum} Our evaluation further confirms that Interlude is capable of real-world deployment. Thus, in our work we present a blockchain protocol designed to work readily in a more practical setting.

\subsection*{Future Work}
\subsubsection{Smart Contracts}
Our approach to split transactions into disjoint sets is limited to cryptocurrencies that only exchange financial value. Recently, blockchains have been used to serve a variety of purposes via the use of ``smart contracts.'' However, techniques required to split smart contracts would need to be more complicated to handle various scenarios that may arise and require further study.

\subsubsection{Solving Network Fairness Issues}
The ideas we propose to solve network fairness, sub-chains that divide the transaction set into disjoint subsets and the Umbrella PoW scheme may be of independent interest to blockchain protocol designers that wish to tackle these issues.

\subsubsection{Need For Game-Theoretic Analysis in Blockchain Protocols}
We re-emphasize the need for blockchain protocol designers to focus on practical aspects such as game-theoretic issues and network fairness concerns which may arise in real-world settings to develop viable solutions for the blockchains of the future.
\newpage
\bibliographystyle{IEEEtran}
\bibliography{references}
\newpage
\appendices
\section{Random Oracle RO\label{appendix_ro}}
\begin{figure}[!htb]
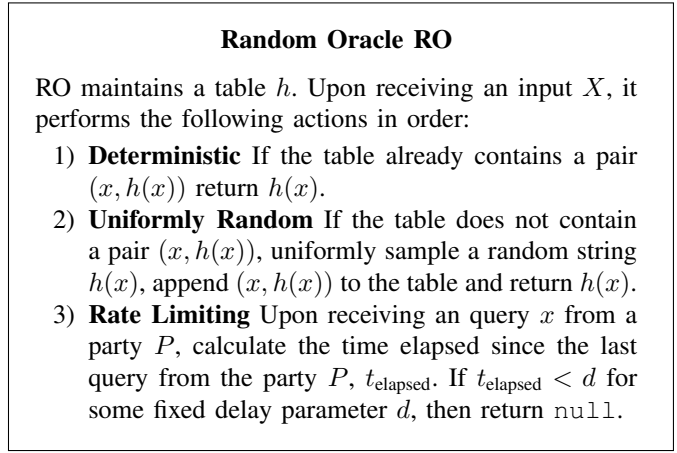

    \centering
    \setlength{\fboxsep}{10pt}%
    \fbox{%
    \begin{minipage}{8cm}
    \begin{center}
        \textbf{\hl{Random Oracle $\text{RO}$}}
    \end{center}
    \hl{RO maintains a table $h$. Upon receiving an input $X$, it performs the following actions in order:}
    \begin{enumerate}
        \item \textbf{Deterministic} If the table already contains a pair $(x,h(x))$ return $h(x)$.
        \item \textbf{Uniformly Random} If the table does not contain a pair $(x,h(x))$, uniformly sample a random string $h(x)$, append $(x, h(x))$ to the table and return $h(x)$.
        \item \textbf{Rate Limiting} Upon receiving an query $x$ from a party $P$, calculate the time elapsed since the last query from the party $P$, $t_\text{elapsed}$. If $t_\text{elapsed} < d$  for some fixed delay parameter $d$, then return \texttt{null}.
    \end{enumerate}
    \end{minipage}
    }
    \caption{Random Oracle $RO$ with parameter $d$\label{random_oracle}}
\end{figure}
\section{Notation\label{notation_table}}
\begin{table}[h!]
\centering
\begin{tabular}{cp{0.8\linewidth}} 
 \toprule
 \textbf{Symbol} & \textbf{Description} \\
 \midrule
 $k$ & Number of chains \\
 $n$ & Number of honest parties \\
 $m$ & Number of adversarial parties \\
 $\lambda$ & Rate parameter of the Poisson process for parallel blocks \\
 $\beta$ & Rate parameter of the Poisson process for series blocks \\ 
 $\delta$ & Network Delay \\
 $\alpha$ & The quantity $m/n$\\
 $\kappa$ & Security parameter chosen by an honest node\\
 $\operatorname{D}(k, l)$ & Erlang Distribution with shape parameter $k$ and rate $l$\\
 $\epsilon$ & The quantity $(1-e^{-\beta\delta})/e^{-\beta\delta}$\\
 $\mathcal{C}$ & The set of all chains\\
 $|C|$ & Number of series blocks in the chain $C$\\
 $C^*$ & The longest chain containing the transaction $\mathcal{T}$ or \mbox{$C^* = \max_{\{C \in \mathcal{C}_i| \mathcal{T} \in C\}} |C|$}\\
 $C'$ & The longest chain not containing the transaction $\mathcal{T}$ or \mbox{$C' = \max_{\{C \in \mathcal{C}_i| \mathcal{T} \notin C\}} |C|$}\\
 \bottomrule
\end{tabular}
\caption{List of symbols used in the paper}
\label{table:1}
\end{table}
\section{Proofs for Lemmas and Theorems in Section~\ref{security_analysis}\label{allproofs}}
\subsection{Proof of Lemma~\ref{eq_fork}\label{eq_fork_appendix}}
\begin{proof}
Let us assume that there are two public forks with a different number of blocks. If an honest party receives both the forks, the fork with the greater number of blocks would be selected via the Max-valid algorithm. However, by definition, a public fork must compete equally with the other forks. So the fork with a lesser number of blocks cannot be a public fork.
Hence, we reach a contradiction.
\end{proof}
\subsection{Proof of Lemma \ref{moreThanf}\label{proof_appendix}}
\begin{proof}
Let us assume that in the worst case, the forks are not resolved in the duration of the round. At the end of the round, there are two possible cases:\\
\noindent \textbf{Case 1:} The honest parties mine a single series block (Figure \ref{fig:lemma})\\
In this case, the adversary would be required to mine atleast $f-1$ series blocks. It would be optimal for the adversary to mine the minimum number of parallel blocks. Hence, the adversary's best strategy would be to use $h$ blocks mined from a chain by the honest parties and append $k-h$ blocks of its own for some $h$. Then the adversary can mine $f-1$ series blocks based on the same set of $k$ parallel blocks.
Let $t_\mathcal{A}$ be the time taken by the adversary to mine $k-h$ blocks. The honest nodes should have mined atleast $h$ blocks by this time. Let us denote $t_\mathcal{H}$ and $t_\mathcal{H}'$ the time taken to mine $f h$ blocks and $f k$ blocks respectively by the honest nodes. (Since, the forks are not resolved between the rounds and the Lemma \ref{eq_fork} holds)

\begin{figure}[!htb]
    \centering
    \includegraphics[width=\linewidth]{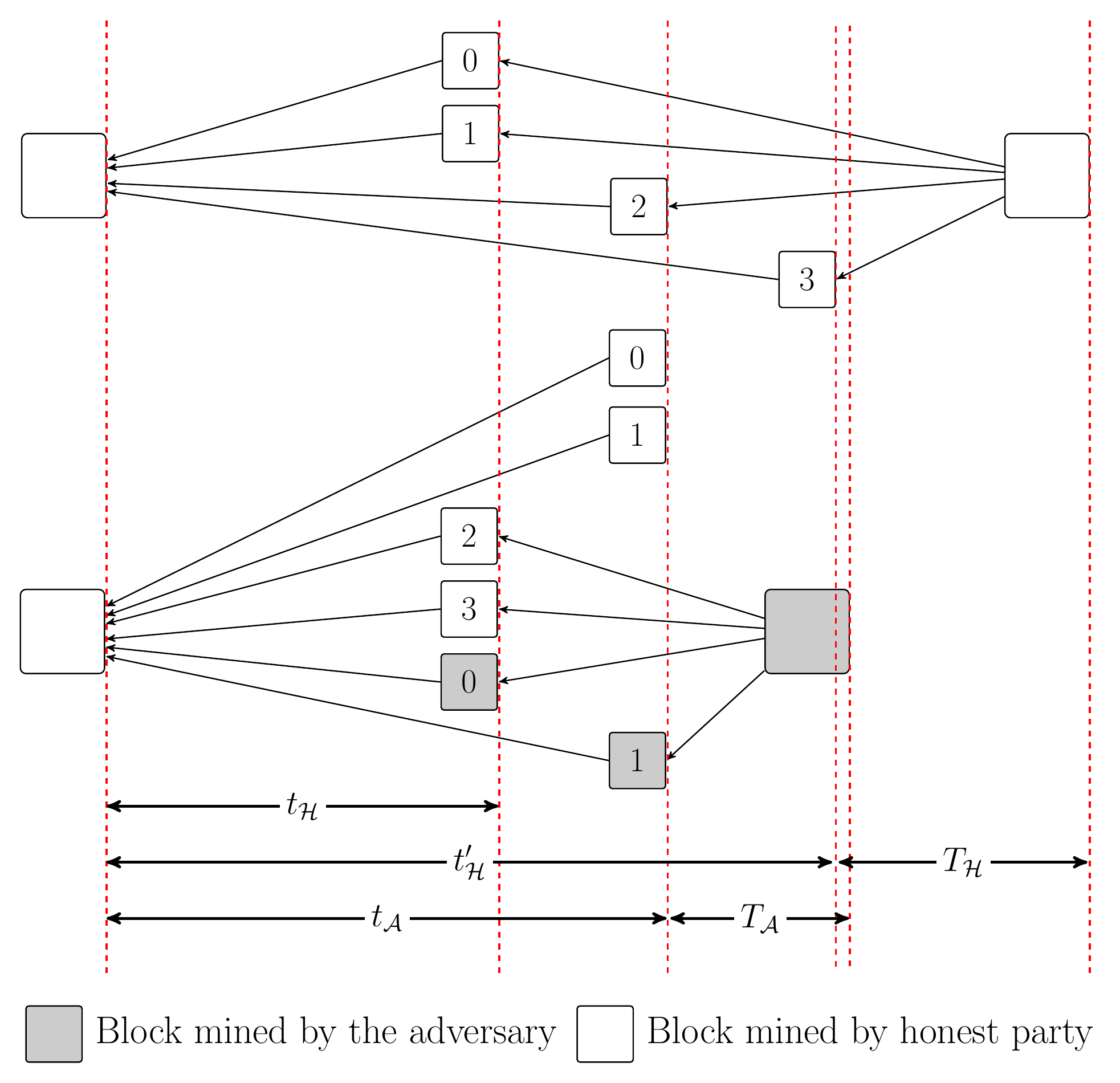}
    \caption{An illustration of two public forks being advanced by a round with $k = 4$ and $h = 2$}
    \label{fig:lemma}
\end{figure}
Let $T_\mathcal{A}$ be the time taken by the adversary to mine $f-1$ series blocks and $T_\mathcal{H}$ be the time taken by the honest parties to mine their first series block. 

In order for the adversary to succeed in this case, inequalities \ref{useHBlocks} and \ref{winRace} must hold true:
\begin{equation}
    \label{useHBlocks}
    t_\mathcal{H} \leq t_\mathcal{A}
\end{equation}
\begin{equation}
    \label{winRace}
    t_\mathcal{A} + T_\mathcal{A} \leq t_\mathcal{H}' + T_\mathcal{H} + \delta
\end{equation}
The random variables follow the following distributions:
\begin{itemize}
    \item $t_\mathcal{H} \sim \operatorname{D}(f h, \lambda)$
    \item $t_\mathcal{A} \sim \operatorname{D}(k-h, \alpha\lambda)$
    \item $t_\mathcal{H}' \sim \operatorname{D}(f k, \lambda)$
    \item $T_\mathcal{H} \sim \operatorname{D}(1, \beta)$
    \item $T_\mathcal{A} \sim \operatorname{D}(f-1, \alpha\beta)$
\end{itemize}
Since it is difficult to obtain a closed-form expression of the above inequality, one can observe via plots that under the assumptions on system parameters,  $\operatorname{P}(t_\mathcal{H} \leq t_\mathcal{A}) \leq 1/2$ for $h \geq k/(\alpha f + 1)$ and $\operatorname{P}(t_\mathcal{A} + T_\mathcal{A} \leq t_\mathcal{H}' + T_\mathcal{H} + \delta) \leq 1/2$ for $h \leq k/(\alpha f + 1)$. The latter turns out to be a decreasing function in both $f$ and $k$, which draws from the intuition that it becomes increasingly difficult for the party to mine $f-1$ serial blocks in the given time. Hence, for all $h$, $\operatorname{P}(t_\mathcal{H} \leq t_\mathcal{A}) \times \operatorname{P}(t_\mathcal{A} + T_\mathcal{A} \leq t_\mathcal{H}' + T_\mathcal{H} + \delta) \leq 1/2$. Thus, the probability that the lemma fails in this case is $1/2$.\\
\noindent \textbf{Case 2:} The honest parties mine two or more than two series blocks\\
This can only happen if the time difference between the mining of the first two blocks is less than $\delta$. Since the mining of blocks is assumed to be a Poisson Process, the probability of such an event is at most $e^{-\beta\delta}$.

Hence, the probability that the conditions of the lemma holds true is at most $1/2+\epsilon$.
\end{proof}
\subsection{Proof of Lemma~\ref{morethanf_1}\label{proof_morethanf_1}}
\begin{proof}
By dividing into two cases as in the proof of Lemma \ref{moreThanf} and proceeding similarly with  $T_\mathcal{A} \sim \operatorname{D}(f, \alpha\beta)$, one can observe that the probability for the first case is bounded by $1/4$ while the probability of the second case remains the same.
\end{proof}
\subsection{Proof of Lemma~\ref{private_difficulty}\label{proof_private_difficulty}}
\begin{proof}
    We require atleast $k$ parallel blocks to advance a private fork by one round while a public fork requires $k-h$ blocks where $h$ blocks come from honest parties. Both require atleast one series block to be mined.
    Hence, if the adversary can accumulate enough PoW to advance the private fork by one round, he/she can also advance a public fork.
\end{proof}
\section{Chain Selection Algorithm\label{chain_selection_code}}
\begin{algorithm}[htbp]
\caption{An algorithm that unanimously picks the longest valid chain}
\label{max-valid}
\hspace*{\algorithmicindent} \textbf{Input}\\
$\mathbf{C}_1, \ldots, \mathbf{C}_m$ - List of chains available\\
\hspace*{\algorithmicindent} \textbf{Output}\\
$\mathbf{C}_\text{max}$ - Longest valid chain
\begin{algorithmic}
\Procedure{MAX-VALID}{$\mathbf{C}_1, \ldots, \mathbf{C}_m$}
\State $\mathbf{C}_\text{max} \gets \epsilon$
\For{$i = 1, m$}
    \If{VALIDATE($\mathbf{C}_i$)}
        \If{$|\mathbf{C}_i| > |\mathbf{C}_\text{max}|$}
            \State $\mathbf{C}_\text{max} \gets \textbf{C}_i$
        \ElsIf{$|\mathbf{C}_i| = |\mathbf{C}_\text{max}|$}
            \State $S_i \gets \text{last $k$ - set of}\ \mathbf{C}_i$
            \State $S_\text{max} \gets \text{last $k$ - set of}\ \mathbf{C}_\text{max}$
            \If{$|S_i| > |S_\text{max}|$}
                \State $\mathbf{C}_\text{max} \gets \textbf{C}_i$
            \ElsIf{$|S_i| = |S_\text{max}|$}
                \State $B_i \gets \text{last series block of}\ \mathbf{C}_i$
                \State $S_\text{max} \gets \text{last series block of}\ \mathbf{C}_\text{max}$
                \If{$B_i.h < B_\text{max}.h$}
                    \State $\mathbf{C}_\text{max} \gets \textbf{C}_i$
                \ElsIf{$B_i = B_\text{max}$}
                    \State{$p \gets \min_{b \in S_i} b.h$}
                    \State{$q \gets \min_{b \in S_\text{max}} b.h$}
                    \If{$p < q$}
                        \State $\mathbf{C}_\text{max} \gets \textbf{C}_i$
                    \EndIf
                \EndIf
            \EndIf
        \EndIf
    \EndIf
\EndFor
\EndProcedure
\end{algorithmic}
\end{algorithm}
\FloatBarrier
\section{Adversarial Strategy\label{adversarial_strategy}}
In this section, we describe a strategy that a potential adversary could adopt in order to disrupt the operation of the distributed ledger.

\cite{sapirshtein2016optimal} described an optimal selfish-mining strategy for linear blockchain protocols such as Bitcoin, we generalize the same for \proto. In our version of the adversarial strategy, if the adversary's private fork is shorter than a public fork, it will fork the longest public fork otherwise it will mine on top of its private fork. In case the public fork advances by a block, the adversary will quickly release another block from its private fork to split the honest mining power.
\end{document}